\documentclass[a4paper,11pt,dvipsnames]{article}
\pdfoutput=1 %
\usepackage{jheppub} % 
\RequirePackage{graphicx}
\usepackage[export]{adjustbox}
\usepackage{amsfonts,amsmath,amssymb}
\usepackage[utf8]{inputenc}
\usepackage{comment}
\usepackage[caption=false]{subfig}
\usepackage{braket}
\usepackage{slashed}
\usepackage[dvipsnames]{xcolor}

\renewcommand{\Im}{\, \mathrm{Im} \,}
\makeatletter
\def\fmslash{\@ifnextchar[{\fmsl@sh}{\fmsl@sh[0mu]}}
\def\fmsl@sh[#1]#2{%
	\mathchoice
	{\@fmsl@sh\displaystyle{#1}{#2}}%
	{\@fmsl@sh\textstyle{#1}{#2}}%
	{\@fmsl@sh\scriptstyle{#1}{#2}}%
	{\@fmsl@sh\scriptscriptstyle{#1}{#2}}}
\def\@fmsl@sh#1#2#3{\m@th\ooalign{$\hfil#1\mkern#2/\hfil$\crcr$#1#3$}}
\makeatother

\begin{document}
\preprint{\parbox[t]{200pt}{SI-HEP-2019-07, P3H-19-34, \\TPP19-032, TUM-HEP-1228/19 \\INT-PUB-19-049 }}

\title{\boldmath The Heavy Quark Expansion for Inclusive Semileptonic Charm Decays Revisited }

\author[a,b]{Matteo Fael,}
\emailAdd{matteo.fael@kit.edu}

\author[a]{Thomas Mannel}
\emailAdd{mannel@physik.uni-siegen.de}

\author[a,c]{and K. Keri Vos}
\emailAdd{keri.vos@tum.de}
\affiliation[a]{Theoretische Physik I, Universit\"at Siegen, Walter-Flex-Strasse 3, 57068 Siegen, Germany}
\affiliation[b]{Institut f\"ur Theoretische Teilchenphysik, Karlsruhe Institute of Technology (KIT), Wolfgang-Gaede Straße 1, 76131 Karlsruhe, Germany}
\affiliation[c]{Physik Department, Technische Universität München, James-Franck-Str. 1, 85748 Garching, Germany}

\abstract{
  The Heavy Quark Expansion (HQE) has become an extremely powerful tool in flavor physics. For charm decays, where the expansion parameters $\alpha_s(m_c)$ and $\Lambda_{\rm QCD}/m_c$ are bigger than for bottom decays, it remains to be seen if the HQE can be applied with similar success. Nevertheless, to make optimal use of the plethora of data already available and coming in the near future, a better understanding of HQE for charm decays is crucial. This paper discusses in detail how the HQE for charm decays is set up, what is the role of four-quark (weak annihilation) operators and how this compares to the well understood bottom decays. Subtleties concerning radiative corrections and the charm mass scheme are briefly discussed. An experimental study of the relevant HQE hadronic matrix elements will then show if the HQE expansion for charm converges well enough. Besides serving as an important cross check for inclusive $B$ decays, in the end, this study might open the road for inclusive $|V_{cs}|$ and $|V_{cd}|$ extractions. 
}

\maketitle
\flushbottom

\section{Introduction}
\label{sec:intro}

Precision calculations for $B$ meson decays rely heavily on the Heavy Quark Expansion (HQE), which makes use of the fact that for sufficiently heavy quarks, the various observables can be expressed as a double expansion in $\alpha_s (m_Q)$ as well as in $\Lambda_{\rm QCD} / m_Q$, where $m_Q$ is the mass of the heavy quark. For inclusive semileptonic $b \to c \ell \nu$ transitions the HQE has become quite sophisticated and high orders in both expansion parameters have been studied~\cite{Jezabek:1988iv,Aquila:2005hq,Pak:2008cp,Melnikov:2008qs,Biswas:2009rb,Becher:2007tk,Gambino:2011cq,Alberti:2013kxa,Mannel:2014xza,Mannel:2019qel,Dassinger:2006md,Mannel:2010wj,Mannel:2018mqv,Fael:2018vsp}. 

In combination with large data samples from CDF, CLEO, DELPHI and $B$ factories~\cite{Acosta:2005qh,Csorna:2004kp,Abdallah:2005cx,Aubert:2004td,Aubert:2009qda, Urquijo:2006wd,Schwanda:2006nf,Bornheim:2002du,Limosani:2005pi,Aubert:2005mg,Aubert:2005im,Bizjak:2005hn,Lees:2011fv}, the HQE allows to determine the CKM parameters $|V_{ub}|$ and $|V_{cb}|$ with a precision of about 6\% and 2\%, respectively~\cite{Gambino:2013rza,Alberti:2014yda,Gambino:2016jkc,Gambino:2007rp,Lange:2005yw,Andersen:2005mj}. 
 
For charm decays, however, one may wonder if the HQE can be applied with similar success. Clearly, the expansion parameters $\alpha_s (m_c)$ and $\Lambda_{\rm QCD} / m_c$ are still less than unity, however they are not particularly small, and hence the HQE cannot be expected to converge as fast as for the bottom quark.
This vice can be turned into a virtue: since $\Lambda_{\rm QCD} / m_c$  is not so small, charm decays are more sensitive to higher-order terms in the HQE than bottom decays. Inclusive charm decays may therefore serve as a tool to study the anatomy of the subleading terms of the HQE. Such a study requires a more detailed understanding of the HQE in charm decays, to which this paper aims. 
%However, in order to get sufficiently precise extractions for the HQE parameters, the HQE for charm decays needs to be understood in more detail.

One of the main challenges in setting up the HQE for charm is that the charm mass is dangerously close to the region where QCD is no longer perturbative. Possible violations of the quark-hadron duality might actually lead to a general failure of the operator product expansion (OPE) at a scale as low as $m_c$.
This can be seen, for instance, in the ground-state charmed-hadron lifetimes which are predicted to be identical in the heavy quark limit, while data show that the $D^\pm$ lifetime is about two and a half times the $D^0$ lifetime. On the contrary, for hadrons containing a bottom quark (but no charm) all lifetimes are equal within a 10\% range. These differences are due to four-quark operators in the HQE, the weak annihilation (WA) and Pauli interference contributions. These operators are sensitive to the flavour of the spectator quark.
Although these terms are formally suppressed by three powers of $m_Q$, they are numerically enhanced by a factor $16 \pi^2$ and they account for the bulk of lifetimes differences in charm decays~\cite{Lenz:2013aua,Lenz:2015dra}.

For inclusive semileptonic $D$ decays the situation seems to be better since the widths of the various charmed hadrons are found to be quite similar~\cite{Asner:2009pu}:
\begin{align}
  \Gamma(D^+ \to X e^+ \nu_e)/\Gamma(D^0 \to X e^+ \nu_e) &=
  0.985 \pm 0.015 \pm 0.024 \, , \notag \\
  \Gamma(D^+_s \to X e^+ \nu_e)/\Gamma(D^0 \to X e^+ \nu_e) &=
  0.828 \pm 0.051 \pm 0.025 \, . \label{eqn:lifedif}
\end{align}
The validity of the HQE for these decays was already studied in the 1990s~\cite{Falk:1995kn,Voloshin:2002je}.
The effect of WA operators was studied in detail in Ref.~\cite{Bigi:1993bh}. Here it was discussed that the WA contribution is concentrated at the end point of the lepton spectrum $q^2 = M_B^2$. The WA contribution is more pronounced in charm decays, therefore CLEO data of the inclusive semileptonic charm decays \cite{Asner:2009pu} were used to determine the size of the WA operators \cite{Ligeti:2010vd, Gambino:2010jz}. The effects of the WA operators was found to be small, which gives confidence in the validity of the HQE for charm decays. On the other hand, in the quest for the highest precision these effects have to be studied in more detail as they contribute both to inclusive $B \to X_u \ell \nu$~\cite{Becirevic:2008us,Bigi:2009ym,Ligeti:2010vd,Gambino:2010jz} and $B\to X_{s,d} \ell\ell$ decays~\cite{Huber:2015sra,Huber:2019iqf}. Therefore, it is important to further constrain the size of WA and the uncertainty associated to it, which requires precise measurements of inclusive $c \to s \ell \nu$ and $c \to d \ell \nu$ transitions.

In charm, there are impressive data sets available and coming up in the near future. Both Belle II and BES III have specific experimental programs dedicated to leptonic and semileptonic $D$ meson decays~\cite{Kou:2018nap,Asner:2008nq}. Moreover, two new Super Tau-Charm Factories, have been proposed at Novosibirsk BINP~\cite{stcfBINP}, Russia, and Hefei USTC~\cite{Luo:2017cwr}, China, to study charm physics in $e^+ e^-$ collisions close to the $D\bar{D}$ threshold with high statistics. In view of this wealth of experimental data, the successful application of the HQE to semileptonic $B$ decays and the hints of its applicability also to $D$, a detailed reanalysis of the HQE for charm is timely and crucial to exploit the full data set. 
Besides, providing information on the non-perturbative HQE elements and the link to $B$ decays, such a study may also open the road to an inclusive measurement of $|V_{cs}|$ and $|V_{cd}|$.  
Theoretically, the two main challenges are: to understand the anatomy of the non-perturbative power corrections in the HQE at higher order and the inclusion of higher-order terms in the $\alpha_s$ expansion both in the partonic rate and the subleading $1/m_c$ corrections. Also, and related to this, the proper choice of the mass scheme for charm is a subtlety that must be addressed. The purpose of the present paper is to set up the HQE for charm decays as an expansion in three parameters, which are $\alpha_s (m_c)$, $\Lambda_{\rm QCD} / m_c$ and $m_s / m_c$, instead of two as in the $b \to c $ case.
By making use of the method of regions~\cite{Beneke:1997zp,Smirnov:2002pj}, we explicitly construct the OPE up to and including terms of order $\Lambda_{\rm QCD}^4 / m_c^4$ and $(m_s / m_c)^4$. 
We will show that compared to the case of $b \to c \ell \bar{\nu}$, this expansion allow us to systematically take into account the additional power corrections given by hadronic matrix elements of four-quark operators. These matrix elements should then be extracted from $q^2$ and energy moments of inclusive semileptonic $D$ decays. Finally, after such a future extraction of the power-suppressed matrix elements, their sizes will indicate if the HQE for charm decays works well enough.  

The paper is organized as follows: we first discuss in section~\ref{sec:HQEcharm} four different cases for setting up the HQE and then we focus on the $c\to s$ transition. The $c\to d$ transition can be trivially obtained from our results, as we discuss later. We subsequently discuss in section~\ref{sec:msexpansion}  the method of regions and the $m_s/m_c$ expansion, the perturbative matching and the mixing of operators in sections~\ref{sec:matching} and~\ref{sec:RGE}. In section~\ref{sec:newHQE}, we give our new HQE matrix elements that should be extracted from data. We discuss some subtleties concerning   QCD corrections and the link between HQE from $B$ and $D$ decays in sections~\ref{sec:QCD} and~\ref{sec:BvsD}. We end with a short outlook and conclusion. 

%%%%%%%%%%%%%%%%%%%%%%%%%%%%%%%%%%%%%%%%%%%%%%%%%%
\section{The Heavy Quark Expansion for Charm}
\label{sec:HQEcharm}
%%%%%%%%%%%%%%%%%%%%%%%%%%%%%%%%%%%%%%%%%%%%%%%%%%
The HQEs for the charm and bottom quark are fundamentally different due to the hierarchy between the mass of the heavy quark $Q$ in the initial state and the  quark $q$ in the final state.
We distinguish four cases:  
\begin{description}
\item[\boldmath I: $m_Q \sim m_q \gg \Lambda_\mathrm{QCD}$] 
This is the usual point of view adopted in the OPE for $b \to c \ell \bar{\nu}$ decays and for the determination of $|V_{cb}|$. 
The quark $q$ is treated as a heavy degree of freedom and therefore the operators arising at tree level are two-quarks operators of the form $\bar{Q}_v (iD^{\mu_1} \dots iD^{\mu_n}) Q_v$ containing only gluons and the quasi-static field $Q_v(x) = \exp(i m_Q v \cdot x) Q(x)$.\footnote{Also four-quark operator with quarks lighter than $q$ can appear.} 
The ratio $m_q / m_Q$, which is assumed to be of order one, appears in the Wilson coefficients of the OPE.
Starting at order $1/m_Q^3$, the HQE develops an infrared sensitivity to the mass of the quark $q$ in the form of a bare logarithm $\log(m_q/m_Q)$ --- there are also power-like singularities like $1/m_q^2$ terms starting from order $1/m_Q^5$~\cite{Mannel:2010wj}.
\item[\boldmath II: $m_Q \gg m_q \gg \Lambda_\mathrm{QCD}$]
In this case it is convenient to first set up an OPE at a scale $\mu \sim m_Q$ where the quark $q$ is still a dynamical degree of freedom.
Four-quark operators of the form $(\bar{Q}_v \Gamma q) (q \bar \Gamma Q_v)$ then appear in this expansion.
After that, the Wilson coefficients are scaled down to $\mu \sim m_q \gg \Lambda_\mathrm{QCD}$  via the renormalization group equation (RGE), where the quark $q$ decouples. Then, we perform a second matching, this time only onto two-quark operators, so that four-quark operators involving the quark $q$ are removed.
Compared to case I, the $\log(m_q/m_Q)$ are produced as an RGE effect.

\item[\boldmath III: $m_Q \gg m_q \sim \Lambda_\mathrm{QCD}$]
  Now, the light quark $q$ remains a dynamical degree of freedom and cannot be integrated out, therefore four-quark operators containing $q$ remain in the OPE. 
The infrared sensitivity to the light degrees of freedom appears as additional non-perturbative parameters, which first appear at $1/m_Q^3$.
The non-analytic term $\log(m_q/m_Q)$, which arises in case I and II, does not explicitly appear as it is hidden inside hadronic matrix elements of four-quark operators of the form $\bra{H} (\bar{Q} \Gamma q) \, (\bar{q} \Gamma^\dagger Q) \ket{H}$. We show that these operators can be absorbed into new non-perturbative HQE parameters. 

\item[\boldmath IV: $m_Q \gg \Lambda_\mathrm{QCD} \gg  m_q$]
This case applies to $b \to u$ and $c \to d$ transitions, since the up and down quark can safely be considered massless. In fact, this case is related to case III by taking the massless limit. 
\end{description}
We focus on the $c\to s$ transition, which has $m_s \sim \Lambda_{\rm QCD}$ and falls into case III. Therefore, compared to $b \to c$, we have a third expansion parameter, $m_s / m_c \sim 1/12$, which has to be treated to be of the same order as  $\Lambda_{\rm QCD} / m_c$ in the HQE. The expansion in $m_s/m_c$ allows us to systematically determine the four-quark operator contributions to total rate and spectral moments and at the same time to establish order by order in the HQE their connection to the two-quark operators via the renormalization group evolution.
%Important here is to understand the origin of the four-quark operator, in particular their connection with the renormalization group evolution.
To this end, we will perform the OPE directly on the expressions for these observables rather than on the differential rate --- along the same lines as in ref.~\cite{Bauer:1996ma} --- therefore after phase space integration.
Our aim is to determine the power corrections to the total rate $\Gamma$ and the moments of kinematical distributions $\langle M^{(n)}[w] \rangle$,\footnote{For the definition see \eqref{eqn:Mndef}.} in terms of a common set of HQE parameters that we denote with $X_i$.

The total width for inclusive semileptonic $D$ meson decay is determined  
as the imaginary part of the forward scattering amplitude~\cite{Shifman:1984wx,Chay:1990da,Manohar:1993qn}
\begin{equation}
  \Gamma = \frac{1}{M_D}  \mathrm{Im} \,
  \bra{D} i \int d^4x \, e^{-ip_D \cdot x} T \left\lbrace \mathcal{H}_W^{\dagger}(x) , \mathcal{H}_W(0) \right\rbrace
  \ket{D} \, =
  \frac{1}{M_D} \, \mathrm{Im} \,
  \bra{D} R \ket{D},
  \label{eqn:Gamma}
\end{equation}
where the weak Hamiltonian is
\begin{equation}
  \mathcal{H}_W = \frac{4 G_F}{\sqrt{2}} V_\mathrm{CKM}
  \left( \bar{q} \gamma^\mu_L c \right)
  \left( \bar{\nu_\ell} \gamma_{L \mu} \ell \right) =
  \frac{4 G_F}{\sqrt{2}} \, V_\mathrm{CKM} \, J^\mu_q \, J_{\ell \mu},
  \label{eqn:HW}
\end{equation}
with $\gamma^\mu_L = \gamma^\mu P_L$, $P_L=(1-\gamma_5)/2$ the left-handed projector, $G_F$ the Fermi constant, $V_\mathrm{CKM}$ the relevant element in the CKM matrix, $J_q^\mu$ and $J_\ell^\mu$ the hadronic and the leptonic currents, respectively.
The composite operator $R$ in~\eqref{eqn:Gamma} admits an OPE written in term of local operators:
\begin{equation}
  \mathrm{Im} \, R =
  \Gamma_0 \,\left [ \sum_{i,k} 
    \frac{C_{k}^\mathrm{2q} (\mu)}{m_c^i} \, O_{i+3,k}^{2q}
    + \sum_{i,j} \frac{C_{j}^\mathrm{4q} (\mu)}{m_c^i} \, O_{i+3,j}^{4q}
  \right] \, ,
  \label{eqn:OPER}
\end{equation}
where the superscript 2q and 4q stand for two- and four-quark operators.
We define $\Gamma_0 = G_F^2 m_c^5 |V_\mathrm{CKM}|^2/(192 \pi^3)$, $C_{k}$ are the Wilson coefficients of the operators $O_{n,k}^{2q}$ and $O_{n,j}^{4q}$ of mass dimension $n$. 
The spectral moments can be described using a similar OPE in terms of the same two- and four-quark operators. We postpone their discussion to the next session.

The computation for $c \to q \ell \nu$ decays proceeds in three steps:
\begin{description}
  \item[Step 1: Matching in Perturbation Theory]
The matching consists of the extraction of the Wilson coefficients $C_n(\mu_c)$ at a scale $\mu_c \sim m_c$. As the OPE is a relation among operators, we can determine the $C_{n}$ by calculating on both the left- and right-hand side of eq.~\eqref{eqn:OPER} matrix elements with free quark and gluon states. 
At this stage, we set up a systematic expansion in $m_s/m_c$ employing the method of regions. This expansion produces simple power corrections in $(m_s/m_c)^n$ that match onto two-quark operators of the form $m_s^n  \bar c_v (i D^{\mu_1} \dots iD^{\mu_n}) c_v$. 
Logarithms of the form $\log(\mu/m_s)$ appear as well on the l.h.s.\ of~\eqref{eqn:OPER}. However on the r.h.s.~of~\eqref{eqn:OPER} the same singularities arise from the one-loop matrix elements of four-quark operators, leaving the Wilson coefficients free of any occurrence of~$\log(\mu/m_s)$.

\item[Step 2: Renormalization Group Evolution]
The Wilson coefficients must be evolved to a lower scale $\mu<m_c$  via the computation of the anomalous dimensions and the solution of the Renormalization Group Equation (RGE). We determine the running just at the leading order $\alpha_s^0$. Even if it is rather trivial at this level, it is instructive for understanding the connection between all the $\log(\mu/m_c)$ and the four-quark operators. 

\item[Step 3: Non-Perturbative Regime]
Total rate and spectral moments are then written in terms of a common set of parameters --- denoted generically by $X_i(\mu)$ --- which correspond to non-perturbative matrix elements of the local operators in the OPE:
\begin{equation}
  2M_D X_i (\mu) \equiv \bra{D} O_i \ket{D}\Big\vert_\mu \, .
\end{equation}
Since all $\log(\mu/m_c)$ terms in the coefficients of two-quark operator are generated by the mixing of the four-quark ones,
\begin{equation}
  C^\mathrm{2q}(\mu) 
  = C^\mathrm{2q}(m_c)  
  + \log \left( \frac{\mu}{m_c} \right) \sum_{j} \hat \gamma^T_{ij} C_j^\mathrm{4q} (m_c),
\end{equation}
where $\hat \gamma$ is the anomalous dimension matrix, we can introduce a set of $\mu$-independent parameters $\tau_i$ by combining together four-quark matrix elements with those of two-quark carrying a $\log(\mu/m_c)$ dependence:
\begin{equation}
\tau_i \sim \bra{D} O_i^\mathrm{4q}\ket{D} + \log(\mu^2/m_c^2) \hat \gamma^T_{ij} \bra{D} O^\mathrm{2q}_j \ket{D} \, .
\end{equation}
We will find that just one (three) parameter(s) contributes to the total rate and spectral moments up to $1/m_c^3$ ($1/m_c^4$).
\end{description}

%

%%%%%%%%%%%%%%%%%%%%%%%%%%%%%%%%%%%%%%%%%%%%%%%%%%
\section{Setting up the OPE}
\label{sec:msexpansion}
%%%%%%%%%%%%%%%%%%%%%%%%%%%%%%%%%%%%%%%%%%%%%%%%%%
%
\begin{figure}[htb]
  \centering
  \subfloat[\label{fig:THH0}]
  {\includegraphics[width=0.32\textwidth]{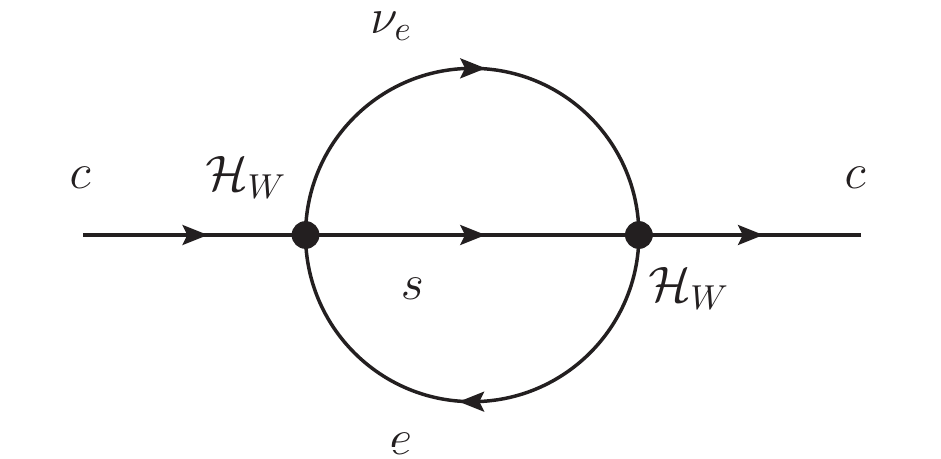}}
  \subfloat[\label{fig:THH1}]{
    \includegraphics[width=0.32\textwidth]{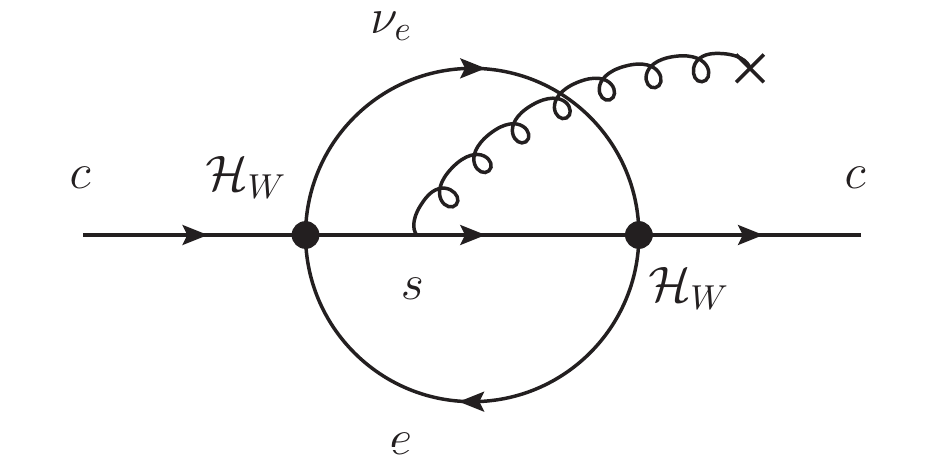}}
  \subfloat[\label{fig:THH2}]{
    \includegraphics[width=0.32\textwidth]{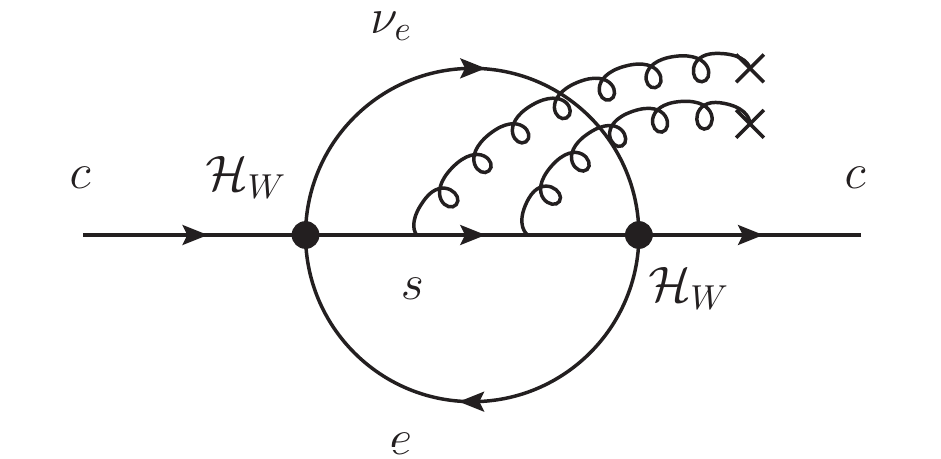}}
    \caption{The zero, one and two gluon matrix elements contributing to the OPE of $\mathrm{Im} \, T \{\mathcal{H}_W^{\dagger} , \mathcal{H}_W \}$.}
  \label{fig:THH}
\end{figure}

We now discuss how the method of regions~\cite{Beneke:1997zp,Smirnov:2002pj} allows us to set up a systematic expansion in $m_s/m_c$. 
In order to perform the matching of Wilson coefficient, we first have to consider in perturbative QCD matrix elements with quarks and gluons on the l.h.s.\ of eq.~\eqref{eqn:OPER}.
Let us consider the diagram~\ref{fig:THH0}, which corresponds to the transition $c \to c$ without any gluon emission.
Its imaginary part gives the rate of $c \to s e \nu$:
\begin{align}
  \bra{c} 2 \Im R \ket{c} &= \int [d^3 p_e] [d^3p_{\nu_e}] [d^3 p_s]
  h_{\mu\nu} (p_s) L^{\mu\nu}(p_e,p_{\nu_e}) \, 
  (2\pi)^4 \, \delta^4(p_c-p_e-p_{\nu_e}-p_s)
  \notag \\
  &=  \int \frac{d q^2}{2\pi} \int [d^3q]  [d^3 p_s]h_{\mu\nu} (p_s)  
   \, (2\pi)^4 \, \delta^4(p_c-q-p_s)  \notag \\
   &\quad \quad \quad \times \int [d^3p_e] [d^3p_{\nu_e}] L^{\mu\nu}(p_e,p_{\nu_e}) \, (2\pi)^4 \delta^4(q-p_e-p_{\nu_e}),
  \label{eqn:PS}
\end{align}
where the momenta of the charm, strange, electron and neutrino are $p_c, p_s, p_e$ and $p_{\nu_e}$, respectively, $q=p_e+p_{\nu_e}$, $[d^3p_i] = \frac{d^3p_i}{(2\pi)^3 2E_i}$, $h_{\mu\nu}(p_s)$ is the hadronic tensor while the leptonic one is
\begin{equation}
  L^{\mu\nu} = 2 \left( 
  p^\mu_e p^\nu_{\nu_e} 
  +p^\nu_e p^\mu_{\nu_e}
  -g^{\mu\nu} p_e \cdot p_{\nu_e}
  \pm i \varepsilon^{\mu\nu\alpha\beta} p_{e \alpha}  \, p_{\nu_e  \beta}
  \right),
\end{equation}
where the upper (lower) sign of the Levi-Civita tensor is for semileptonic bottom (charm) decays. For simplicity we omit an overall constant $8 G_F^2 |V_{cs}|^2$ in eq.~\eqref{eqn:PS}.
Rewriting $[d^3 p_i] = \frac{d^4 p_i}{(2\pi)^4} (2 \pi) \delta_+(p_i^2-m_i^2)$ and integrating w.r.t.\ $d^4p_s$, we can express the total rate as
\begin{multline}
  \bra{c} 2 \Im R \ket{c} 
   =
  \int \frac{dQ^2}{2\pi} 
  \int \frac{d^4 q}{(2\pi)^4} \, 
  (2 \pi) \delta_+((p_c-q)^2-m_s^2) (2\pi) \delta_+(q^2-Q^2) 
  h_{\mu\nu}(p_c-q) \, \mathcal{L}^{\mu\nu}(q)  \\
  = 
  \int \frac{dQ^2}{2\pi} 
  \, 2 \, \mathrm{Im} \, \left[
  \int \frac{d^4 q}{(2\pi)^4} \,
  \bar{u}(p_c) \gamma^\mu_L \frac{i}{\slashed p_c - \slashed q -m_s+i \varepsilon } \gamma^\nu_L u(p_c) \,
  \frac{i}
{q^2 -Q^2 + i \varepsilon} \mathcal{L}_{\mu\nu}(q) \right],
  \label{eqn:Gammaint}
\end{multline}
where we defined the integrated leptonic tensor as
\begin{equation}
  \mathcal{L}^{\mu\nu} (q) =  \int [d^3 p_e] [d^3 p_{\nu_e}]  \,
  L^{\mu\nu}(p_e,p_{\nu_e}) \,
  (2 \pi)^4 \delta^4 (q-p_e-p_{\nu_e}).
\end{equation}
%df
We have written the total rate as the imaginary part of a loop integral containing two massive propagators, $1/(q^2-Q^2)$ and $1/[(p_c-q)^2-m_s^2]$.
We calculate $\bra{c} 2 \Im R \ket{c}$ by dividing the integral in~\eqref{eqn:Gammaint} in two different domains and expanding the integrand into a Taylor series, one where $p_c-q \sim m_c$ (large region) and the other one where $p_c-q \sim m_s$ (small region).

\vspace{0.2cm}
\textbf{The Large Region:} 
The first contribution arises from the region where the loop momentum  in~\eqref{eqn:Gammaint} is large compared to the strange mass, i.e.\ $p_c- q \sim m_c  \gg m_s$. This allows us to Taylor expand the propagator of the strange quark.
At the same time we implement the heavy quark expansion by writing the free charm momentum $p_c = m_c v + k$ which splits the quarks momentum into a large part $ m_c v$ and a residual part with $k \ll m_c$.
The strange propagator can be written as a series
\begin{equation}
  \frac{1}{\slashed p_c - \slashed q-m_s}  = \frac{1}{m_c \slashed v -\slashed q} \sum_{n=0}^\infty \left( (-\slashed k + m_s)\frac{1}{m_c \slashed v -\slashed q} \right)^n.
  \label{eqn:largeexp}
\end{equation}
Actually, the $1/m_c$ expansion in the large region is most conveniently derived following~\cite{Novikov:1983gd,Mannel:2010wj,Dassinger:2006md} by introducing a background field propagator for the intermediate strange quark and expanding it in the following way:
\begin{equation}
S_{\rm BGF}(S) = 
\frac{1}{\slashed S + i\slashed{D} - m_s}  =
\frac{1}{\slashed{S}} 
\sum_{n=0}^\infty  \left( (-i\slashed{D}+m_s) \frac{1}{ \slashed{S}} \right)^n \, , 
\label{eqn:SBGFexp}
\end{equation}
with $S=m_c v-q$.
It yields the usual HQE plus the power corrections in $m_s/m_c$. 
The large region in the end corresponds to the initial phase space integration~\eqref{eqn:PS}, where the light quark in the final state is taken massless.
Since the phase space integral is not finite in four dimensions, we must employ dimensional regularization in its the evaluation. All the $1/\varepsilon$ poles arising from this region will eventually cancel out against those coming from the small region. 

\vspace{0.2cm} 
\textbf{The Small Region:} 
The second contribution comes from the region where the loop momentum $ p_s = p_c-q \sim m_s \ll m_c$, such that the strange propagator must be left unexpanded, while the other one depending on $Q^2$ is  rewritten as (shifting the loop momenta according to $q= p_c-p_s$):
\begin{equation}
  \frac{1}{(p_c-p_s)^2-Q^2}  = \frac{1}{p_c^2-Q^2} \sum_{n=0}^\infty \left( \frac{2p_c \cdot p_s - p_s^2}{p_c^2-Q^2} \right)^n.
  \label{eqn:smallexp}
\end{equation}
The first term in the expansion for~\eqref{eqn:Gammaint} gives
\begin{multline}
  \bra{c} 2 \Im R \ket{c}_\mathrm{small}
  = \\
  \int \frac{dQ^2}{2\pi} 
  \,2 \, \mathrm{Im} \, \left[ \frac{i}{p_c^2-Q^2+i \varepsilon} 
    \int \frac{d^d p_s}{(2\pi)^d} \,
  \bar{u}(p_c) \gamma^\mu_L \frac{i}{\slashed p_s - m_s+i \varepsilon } \gamma^\nu_L u(p_c) \,
\mathcal{L}_{\mu\nu}(p_c-p_s) \right].
\label{eqn:Gammasmallint}
\end{multline}
The imaginary part of \eqref{eqn:Gammasmallint} is given solely by the imaginary part of the $1/(p_c^2-Q^2+i\varepsilon)$ propagator ($- \pi \delta(p_c^2-Q^2)$) because the integral w.r.t.\ $p_s$ correspond to a tadpole diagram and therefore it is real. 
The integration w.r.t.\ $Q^2$ yields the condition $Q^2=m_c^2$. 
Note in addition that higher order terms in the series~\eqref{eqn:smallexp} do not contribute since 
\begin{equation}
  \mathrm{Im}  \left( \frac{1}{p_c^2-Q^2+i\varepsilon} \right)^{n+1} =
  \frac{\pi}{n!} \frac{d^n}{d Q^{2n}} \delta(p_c^2-Q^2) .
\end{equation}
Integrating by parts we would bring the derivatives on the one-loop integral, which however is $Q^2$-independent.  
Eventually the contribution of the small region is:
\begin{equation}
  \bra{c} 2 \Im R \ket{c}_\mathrm{small}
  =
  \int \frac{d^d p_s}{(2\pi)^d} \,
  \bar{u}(p_c) \gamma^\mu_L \frac{1}{\slashed p_s - m_s+i \varepsilon } \gamma^\nu_L u(p_c) \,
\mathcal{L}_{\mu\nu}(p_c-p_s).
\end{equation}
Note that the $\mathcal{L}^{\mu\nu}(q)$ must be evaluated in $d$ dimensions.
The leptonic tensor is transverse for massless leptons, $q_\mu \mathcal{L}^{\mu\nu}=0$, so that its most general form is 
\begin{multline}
  \mathcal{L}^{\mu \nu}(q,v) = 
   \left( q^\mu q^\nu-g^{\mu\nu}q^2 \right)\mathcal{L}_1(q^2, q \cdot v)
   +i\epsilon^{\mu\nu\alpha\beta}v_\alpha q_\beta \;\mathcal{L}_2  (q^2, q \cdot v) \\[5pt]
  +
  \left(v^\mu - q^\mu \frac{v \cdot q}{q^2}\right) 
  \left(v^\nu - q^\nu \frac{v \cdot q}{q^2}\right) \mathcal{L}_3 (q^2, q \cdot v)\, .
  \label{eqn:Lstruct}
\end{multline}
For the total rate and the $q^2$ moments $\mathcal{L}_{2,3}=0$, however for other observables, as e.g.\ the energy moments, the complete structure has to be considered.
Evaluating \eqref{eqn:Lstruct} at $q=p_c-p_s$ and expanding it up to second order in $m_c$, i.e.\ substituting $p_c-p_s = m_c v + t$ with $t \ll m_c$, we find the most general expression for the leptonic tensor in Eq.~\eqref{eqn:Gammasmallint}:
\begin{align}
  \frac{\mathcal{L}^{\mu\nu}(p_c-p_s,v)}{m_c^2} &= 
   \left( 
  v^\mu v^\nu-g^{\mu\nu} +\frac{v^\mu t^\nu+t^\mu v^\nu-2 g^{\mu\nu} v \cdot t}{m_c}
  \right) \mathcal{L}_1(m_c^2, m_c) \notag \\
  &+ \frac{v \cdot t (v^\mu v^\nu-g^{\mu\nu})}{m_c} \mathcal{L}_1'(m_c^2, m_c) 
  + i \epsilon^{\mu\nu\alpha\beta}v_\alpha t_\beta \,  \mathcal{L}_2 (m_c^2, m_c)
  +\dots
 \label{eqn:Lexp}
\end{align}
with $\mathcal{L}_1' = [ 2 \frac{\partial}{\partial q^2} + \frac{\partial}{\partial v\cdot q}]\mathcal{L}_1$ and where the dots represent higher order terms in the $m_c$ expansion.
The structure $\mathcal{L}_3$ starts to contribute only in the sub-sub-leading term proportional to $t^2$.
The expansion~\eqref{eqn:Lexp} allows us to systematically identify each term in $\bra{c} 2 \Im R \ket{c}_\mathrm{small}$ as one-loop matrix elements of four-quark operator $O_i$:  $\bra{c} O_i \ket{c}$. 
WA contributions therefore naturally arise once we set up an $m_s/m_c$ expansion, and they are given by the contraction of the two hadronic currents $J_q^\mu J^{\dagger\nu}_q$ with the leptonic tensor taken at the end point $q^2 = m_c^2$. This fact was discussed already in~\cite{Bigi:1993bh}, however it is based on the analysis of intermediate state saturation of eq.~\eqref{eqn:Gamma} when both the energy and the momentum of the hadronic final state are small compared to $m_c$.

\begin{figure}[h]
  \centering
  \subfloat[\label{fig:mixzero}]{\includegraphics[width=0.24\textwidth]{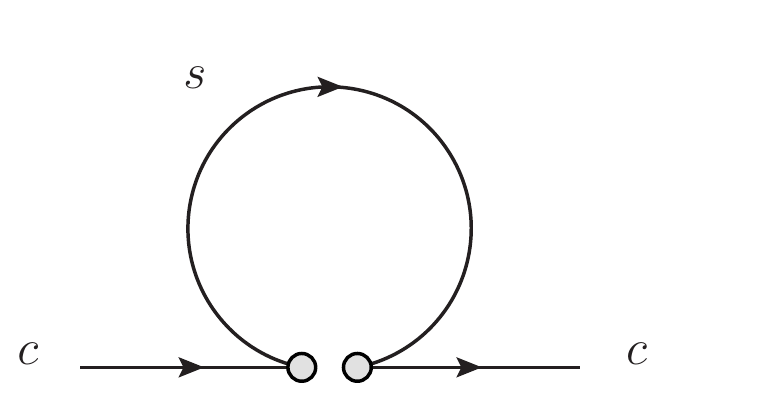}}
  \subfloat[\label{fig:mixone}]{\includegraphics[width=0.24\textwidth]{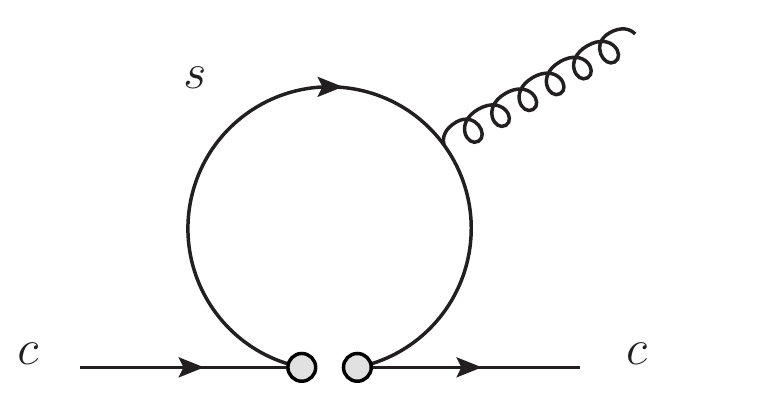}}
  \subfloat[\label{fig:mixtwo}]{
  \includegraphics[width=0.24\textwidth]{m4mix}
  \includegraphics[width=0.24\textwidth]{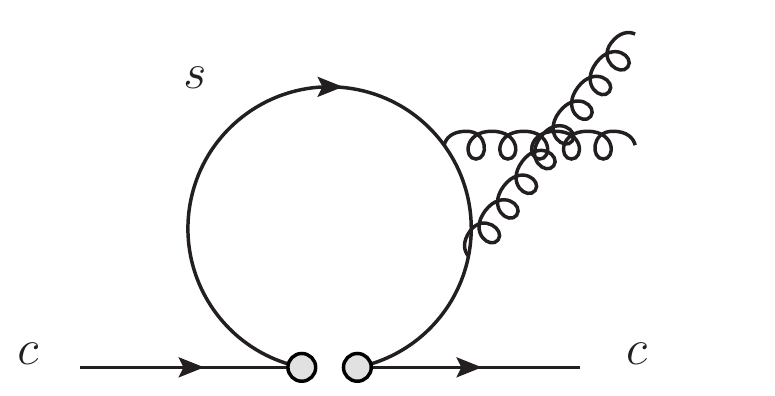}}
  \caption{Diagrams determining the mixing of four-quark operators into two-quark operators.}
  \label{fig:mixing}
\end{figure}

In addition to the zero gluon matrix element  considered so far, we have to take into account for the OPE also matrix elements with soft gluon emission (see figure~\ref{fig:THH}b,c).
Indeed a simple expansion in the residual momentum $k$ yields only the symmetric parts of an operator like $\bar{c}_v (iD^{\mu_1} \dots iD^{\mu_n}) c_v$. 
In order to pin down the antisymmetric part,  we must consider also $c \to c + n$ gluon matrix elements, where the hadronic tensor has the form
\begin{equation}
  \bar{u}(p_c) \gamma^\mu_L \left[
   \frac{1}{\slashed p_s -m_s}
  \slashed \epsilon_1 T^{a_1} \frac{1}{\slashed p_s+\slashed r_1 -m_s} 
  \dots
  \slashed \epsilon_n T^{a_n} \frac{1}{\slashed p_s+ \dots + \slashed r_n -m_s}
+\mathrm{perm.} \right]
  \gamma^\nu_L u(p_c),
  \label{eqn:hgluons}
\end{equation}
with $r_i$, $a_i$ and $\epsilon_i$ the momentum, color index and polarization vector of the $i$-th gluon.
The $m_s$ expansion proceeds along the same line as for the zero-gluon matrix element, keeping in mind that since all the gluons are soft we have $ r_1, \dots, r_n \sim \Lambda_\mathrm{QCD} \ll m_c$.
Also in this case we separate two regions: a large one where each of the propagator in~\eqref{eqn:hgluons} is expanded similarly to eq.~\eqref{eqn:largeexp}. 
The small region requires the expansion of $1/(q^2-Q^2)$. Therefore, $\bra{c+ng} 2 \Im R \ket{c}_\mathrm{small}$ can be seen also as the $n$-gluon matrix elements of four-quark operators $\bra{c+ ng }O_i\ket{c}$ (see figure~\ref{fig:mixone} and~\ref{fig:mixtwo}).

%%%%%%%%%%%%%%%%%%%%%%%%%%%%%%%%%%%%%%%%%%%%%%%%%%
\section{The matching}
\label{sec:matching}
%%%%%%%%%%%%%%%%%%%%%%%%%%%%%%%%%%%%%%%%%%%%%%%%%%
\begin{figure}[htb]
  \centering
  \includegraphics[width=0.32\textwidth]{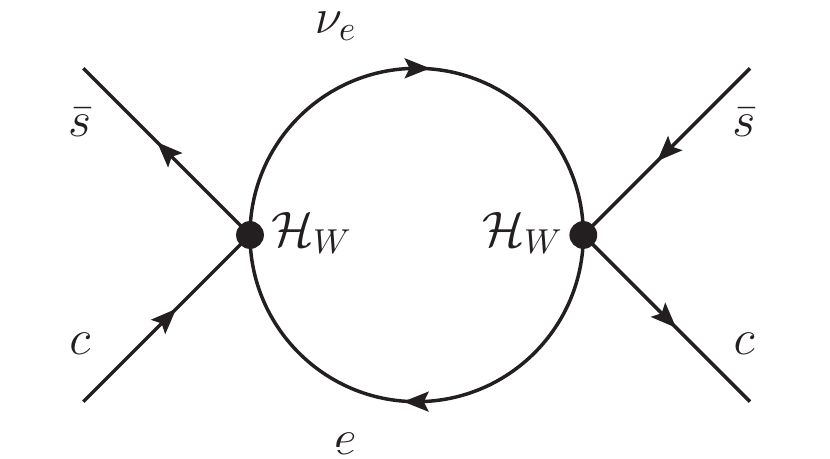}
  \caption{The four-quark matrix element contributing to the OPE of $\Im T \{\mathcal{H}_W^{\dagger} , \mathcal{H}_W \}$.}
  \label{fig:T4Q}
\end{figure}

With the set-up introduced in the previous section, we can now compute the matching conditions for the total rate and the moments of kinematical distributions.
We can define the normalized moments in a generic way as the phase-space integral of the differential rate multiplied by an appropriate weight function $w$~\cite{Fael:2018vsp}:
\begin{equation}
  \langle M^{(n)} [w] \rangle = 
  \frac{1}{\Gamma_0}
  \int d\Phi \,  w^n(v,p_e,p_\nu) \,
  W^{\mu\nu}L_{\mu\nu}.
  \label{eqn:Mndef}
\end{equation}
where $W^{\mu\nu}$ is the hadronic tensor encoding the non perturbative dynamics.\footnote{Actually the ratio $\langle M^{(n)}[w] \rangle/\langle M^{(0)}[w] \rangle$ is measured experimentally. In the following, we concentrate on the OPE for the numerators, as the ratio can be easily obtained from $\langle M^{(n)}[w] \rangle$.}
The weight function $w$ can also contain a phase space cut in the form of Heaviside functions.
The uncut moments of the charged lepton energy $E_e$ and the leptonic invariant mass $q^2$ are given by:
\begin{align}
  \mathcal{Q}^{(n)} &= \frac{1}{\Gamma_0} \int_0^{\hat q^2_\mathrm{max}} (\hat q^2)^n 
  \frac{d \Gamma}{d \hat q^2} d \hat q^2, &
  \mathcal{Y}^{(n)} &= \frac{1}{\Gamma_0} \int_0^{\hat y_\mathrm{max}} 
  \hat y^n \frac{d \Gamma}{d\hat y} d \hat y,
\end{align}
with $\hat q^2 = q^2/m_c^2$ and $y= 2 E_e/m_c$, with the corresponding weight functions $w(v,p_e,p_\nu) = q^{2}/m_c$ and $w(v,p_e,p_\nu) = 2 v \cdot p_e/m_c$, respectively.
Also the spectral moments have an OPE:
\begin{align}
  2 \Im R_{q^2}^{(n)} &= 
   \sum_{i,k} \frac{C_{q^2 \, k}^{(n)}}{m_c^i} \, O_{i+3,k}^{2q}
    + 
    \sum_{i,j} \frac{C_{q^2 \, j}^{(n)}}{m_c^i} \, O_{i+3,j}^{4q} 
  , \notag \\[5pt]
  \label{eqn:OPERM}
  2 \Im R_{E}^{(n)} &= 
    \sum_{i,k} \frac{C_{E \, k}^{(n)}}{m_c^i} \, O_{i+3,k}^{2q}
    + 
    \sum_{i,j} \frac{C_{E \, j}^{(n)}}{m_c^i} \, O_{i+3,j}^{4q} \,.
\end{align}
Here $R_{q^2}^{(n)}$ and $R_{E}^{(n)}$ denote the two composite operators giving rise to the $q^2$ and electron energy moments, respectively. Formally they can be written in terms of the modified Hamiltonian $\mathcal{H}_{q^2} \propto (\bar{q} \gamma^\mu P_L c) \square (\bar{\nu}_\ell \gamma_\mu P_L \ell)$ and  $\mathcal{H}_{E} \propto (\bar{q} \gamma^\mu P_L c) (\bar{\nu}_\ell \gamma_\mu P_L v\cdot \partial \ell)$.
In practice, however the OPE for the moments can be determined by inserting the weight function $w$ into the definition of the leptonic tensor in eq.~\eqref{eqn:Gammaint},
\begin{equation}
  \mathcal{L}^{\mu\nu}_w (q,v) = 
  \frac{1}{\Gamma_0}
  \int [d^3 p_e][d^3p_\nu]  \,
  w^n(v,p_e,p_{\nu_e}) \,
  L^{\mu\nu}(p_e,p_{\nu_e}) \,
  (2\pi)^4 \delta^{(4)} (q-p_e-p_\nu) \, .
  \label{eqn:Lwdef}
\end{equation}
%

%-------------------------------------------
\subsection{Four-quark operators}
\label{sec:4qmatch}
%-------------------------------------------
At tree-level, there are two four-quark operators of dimension six:
\begin{align}
O_1 &=  (\bar{c}_v  \slashed v P_L s) \, (\bar{s} \slashed v P_L c_v), \notag \\
O_2 &=  (\bar{c}_v  \gamma^\mu P_L s) \, (\bar{s} \gamma_\mu P_L c_v). \label{eqn:4qdefdim6}
\end{align}
At dimension seven, there are four:
\begin{align}
  O_3 &= \frac{1}{2} \Big[
    (\bar{c}_v   \gamma^\mu P_L s) (  v\cdot i\partial \, \bar{s} \gamma_\mu P_L c_v)
  -( v\cdot i\partial \, \bar{c}_v   \gamma^\mu P_L s) (\bar{s} \gamma_\mu P_L c_v)
\Big], \notag \\[5pt]
  O_4 &= (\bar{c}_v   \slashed v P_L s) ( i \partial ^\mu \, \bar{s} \gamma_\mu P_L c_v)
  - (i  \partial^\mu \, \bar{c}_v   \gamma_\mu P_L s) (\bar{s} \slashed v P_L c_v),\notag \\[5pt]
  O_5 &= \frac{1}{2} \Big[
    (\bar{c}_v \slashed v P_L s) ( v \cdot i \partial \, \bar{s} \slashed v P_L c_v)
  - ( v \cdot i  \partial \bar{c}_v \slashed v P_L s) (\bar{s} \slashed v P_L c_v)\Big], \notag\\[5pt]
  O_6 &=  \frac{1}{2} (- i \varepsilon_{\mu\nu\rho\alpha} v^\alpha) \Big[ 
  (\bar{c}_v   \gamma^\mu P_L s) (  i\partial^\rho \, \bar{s} \gamma^\nu P_L c_v)
  -( i\partial^\rho \, \bar{c}_v   \gamma^\mu P_L s) (\bar{s} \gamma^\nu P_L c_v)
  \Big],
  \label{eqn:4qdefdim7}
\end{align}
where the derivatives act on both fields inside a bilinear. 
The coefficients $1/2$ are introduced for convenience when calculating Feynman rules.
QCD radiative corrections may induce additional operators where the identity in color space appearing inside the bilinear is substituted with Gell-Mann matrices $T^A$.

We keep the $\slashed v$ inside the definition of the operators, even though it can be rewritten via the equations of motion, as it simplifies the computation of the one-loop matrix elements.
Also, since the coefficients of the four-quark operators are derived from the (transverse) leptonic tensor evaluated near the end point $q^2 \sim m_c^2$~\eqref{eqn:Lexp}, only the following three combinations actually appear in the total rate and the spectral moments up to order $1/m_c^4$:
\begin{align}
  O_0 &= O_1-O_2+\frac{O_4-2O_3}{m_c} \, , \notag\\ 
  O_m &= O_5-O_3 \, ,\notag\\ 
  O_\epsilon &= O_6 \, .
  \label{eqn:defO0ne}
\end{align}
Their Wilson coefficients are computed by considering matrix elements with quark states: $\bra{c\bar{s}} 2 \Im R \ket{c\bar{s}}$.
For the total rate we have
\begin{equation}
  \bra{c\bar{s}} 2 \Im R \ket{c\bar{s}} =
  \Gamma_0 
  \frac{128\pi^2 }{m_c^5} (q^\mu q^\nu-g^{\mu\nu} q^2) 
  \left( \bar{s} \gamma_\mu P_L c \right) \left( \bar{c} \gamma_\nu P_L s \right).
  \label{eqn:4qampmatching}
\end{equation}
By substituting $q = m_c v + t$ and expanding in powers of $m_c$ up to second order as we did in~\eqref{eqn:Lexp},
\begin{equation}
  \bra{c\bar{s}} 2 \Im R \ket{c\bar{s}} =
  \Gamma_0 
  128\pi^2
  \left[ 
    \frac{v_\mu v_\nu-g_{\mu\nu}}{m_c^3} 
    +\frac{v_\mu t_\nu+v_\nu t_\mu-2 \, v \cdot t \, g_{\mu\nu} }{m_c^4}
  \right] 
  \left( \bar{s} \gamma_\mu P_L c \right) \left( \bar{c} \gamma_\nu P_L s \right),
\end{equation}
we find as matching conditions:
\begin{align}
  C_0 &= 128 \pi^2, &
  C_m &= 0  , &
  C_\epsilon & =  0 \, .
\end{align}
To calculate the matching for the $q^2$ moments we use the weight function $w^n(v,p_e,p_\nu) = (q^2/m_c^2)^n$ into~\eqref{eqn:Lwdef}:
\begin{equation}
  \bra{c\bar{s}} 2 \Im R^{(n)}_{q^2} \ket{c\bar{s}} =
  \frac{128\pi^2 }{m_c^5} \, \left(\frac{q^2}{m_c^2}\right)^n (q^\mu q^\nu-g^{\mu\nu} q^2)
  \left( \bar{s} \gamma_\mu P_L c \right) \left( \bar{c} \gamma_\nu P_L s \right).
\end{equation}
The Wilson coefficients for the $q^2$-moments are therefore:
\begin{align}
  C_{q^2,\, 0}^{(n)} &= 128 \pi^2, &
  C_{q^2,\, m}^{(n)} &=  128 \pi^2 \; (2n), &
  C_{q^2,\, \epsilon}^{(n)} &= 0. 
  \label{eqn:matchingq2}
\end{align}
For the charged lepton energy moments we must employ instead $w^n(v,p_e,p_\nu) = (2 p_e \cdot v/m_c)^n $. The leptonic tensor then depends on both $q$ and $v$, however its expression cannot be cast in a simple form for a generic $n$, $q$ and $v$, as for the other two cases. 
Nevertheless, one can substitute in the integrand $q=m_c v+t$ and expand in $t$ up to second order in $m_c$:
\begin{multline}
  \bra{c\bar{s}} 2 \Im R^{(n)}_{E} \ket{c\bar{s}} =
  128\pi^2 
 \left( \bar{s} \gamma^\mu P_L c \right) \left( \bar{c} \gamma^\nu P_L s \right)
  \left[ 
    \frac{v_\mu v_\nu-g_{\mu\nu}}{m_c^3} \right. \\
    \left. + 
     \frac{v_\mu t_\nu + v_\nu t_\mu-(2+n)v\cdot t g_{\mu\nu}
      +n \, v \cdot t \,  v^\mu v^\nu
      \pm i (n/2) \varepsilon_{\mu\nu\alpha\beta} \, t^\alpha v^\beta
    }{m_c^4} + \dots 
  \right] .
\end{multline}
Therefore, the coefficients for the electron energy moments are  
\begin{align}
  C_{E,\, 0}^{(n)} &= 128 \pi^2 , &
  C_{E,\, m}^{(n)} &=  128 \pi^2 \;n, &
  C_{E,\, \epsilon}^{(n)} &= \pm  128 \pi^2 \; \frac{n}{2} ,
  \label{eqn:matchingE}
\end{align}
where the sign in $C_{E,\, \epsilon}$ is plus (minus) for the $c\to s$ ($b \to c$) transition.

%-------------------------------------
\subsection{Two-quark operators}
\label{sec:2qmatch}
\begin{figure}[h]
\begin{multline*}
  C^\mathrm{2q} \times \includegraphics[width=0.15\textwidth,valign=c]{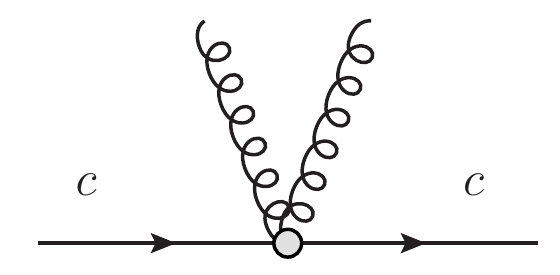} = \\[15pt]
  \includegraphics[width=0.25\textwidth,valign=c]{THH2}\Bigg|_\mathrm{large}
  +\includegraphics[width=0.25\textwidth,valign=c]{THH2}\Bigg|_\mathrm{small} - C^\mathrm{4q} \times \includegraphics[width=0.2\textwidth,valign=c]{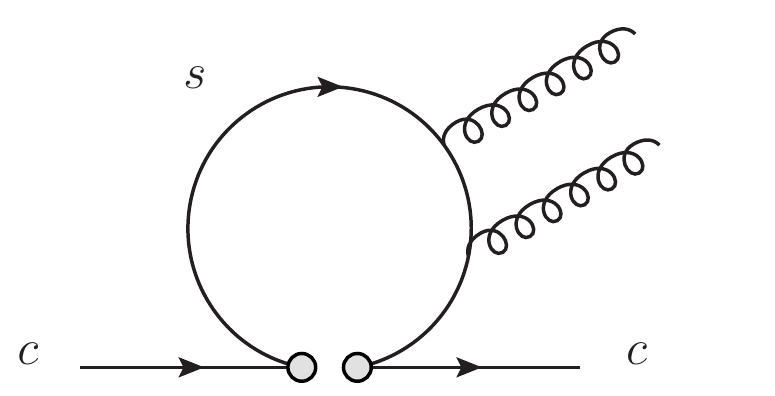}\Bigg|_\mathrm{ren} 
\end{multline*}
\caption{Computation of the Wilson coefficients of two-quark operators. Matrix elements with either zero, one and two gluon must be considered.}
\label{fig:matching2q} 
\end{figure}

%-------------------------------------
The evaluation of the Wilson coefficients of the two-quark operators is more involved. They cannot be determined naively from the known expression for the $b \to c \ell \nu$.
We define the HQE operators up to order $1/m_c^4$ following refs.~\cite{Mannel:2018mqv,Fael:2018vsp}:
\begin{align}
   O_{\mu_3} &= \bar{c}_v c_v, & 
   O_{r_G} &=
   \bar{c}_v \left[ (iD_\mu) \, , \,  (iD_\nu) \right]  \left[ (iD^\mu)  \, , \, (i D^\nu) \right]   c_v \,  , \notag \\[5pt]
   O_{\mu_\pi} &= \bar{c}_v (i D)^2 c_v, & 
  O_{r_E} &= 
  \bar{c}_v \left[ (ivD ) \, , \,  (iD_\mu) \right]  \left[ (ivD)  \, , \, (i D^\mu) \right]   c_v \,  , \notag \\[5pt]
   O_{\mu_G} &= \bar{c}_v \sigma \cdot G c_v,& 
  O_{s_B} &=
  \bar{c}_v \left[ (iD_\mu) \, , \,  (iD_\alpha) \right]  \left[ (iD^\mu)  \, , \, (i D_\beta) \right]   (-i \sigma^{\alpha \beta})  c_v \,  ,
 \notag \\[5pt]
   O_{\rho_D} &= 
   \frac{1}{2} \bar{c}_v \left[ iD^\mu, \left[i v D, i D_\mu \right] \right] c_v  \, ,&
 O_{s_E} &=
 \bar{c}_v \left[ (ivD) \, , \,  (iD_\alpha) \right]  \left[ (ivD)  \, , \, (i D_\beta) \right]   (-i \sigma^{\alpha \beta})  c_v\,   ,
 \notag  \\[5pt]
   O_{\delta \rho_D} &= 
   \frac{1}{2} \bar{c}_v \left[ iD^\mu, \left[(i D)^2, i D_\mu \right] \right] c_v  \, ,&
 O_{s_{qB}} &=
 \bar{c}_v  \left[ iD_\mu \, , \, \left[ iD^\mu \, , \,  \left[ iD_\alpha \, , \, iD_\beta \right] \right] \right]  (-i \sigma^{\alpha \beta}) c_v\,  ,
 \label{eqn:opm4}
\end{align}
with $\sigma \cdot G \equiv -i \sigma^{\mu\nu} (iD_\mu)(iD_\nu)$ and $\sigma^{\mu\nu} = \frac{i}{2}[\gamma^\mu,\gamma^\nu]$.
In addition, we note that
\begin{align}
  O_{\mu_3} &= 1 + \frac{1}{2m_c^2} \left( O_{\mu_G}-O_{\mu_\pi} \right), & 
  O_{\tilde \rho_D} &= O_{\rho_D} +\frac{1}{2m_c}O_{\delta \rho_D}.
 \label{eqn:opm4extra}
\end{align}
Equations~\eqref{eqn:opm4} and~\eqref{eqn:opm4extra} constitute a set of operators which describe observables invariant under a reparametrization transformation $v \to v + \delta v$, such as the total rate and the $q^2$ moments~\cite{Mannel:2018mqv}.
The prediction for non-RPI observables, like the electron energy spectrum, depend at tree-level up to $1/m_c^4$ on a larger set of operators, which include the additional operators:
\begin{align}
  O_{\rho_{LS}} &= \frac{1}{2} \bar{c}_v \left\lbrace iD_\alpha, \left[(i v D), i D_\beta \right] (-i \sigma^{\alpha \beta})  \right\rbrace c_v \, ,
  \notag \\[5pt]
  O_{\delta \rho_{LS}} &= \frac{1}{2} \bar{c}_v \left\lbrace iD_\alpha, \left[(i D)^2, i D_\beta \right] (-i \sigma^{\alpha \beta})  \right\rbrace c_v \, ,
  \notag \\[5pt]
  O_{\delta G_1} &= \bar{c}_v ((i D)^2)^2 c_v ,
  \notag \\[5pt]
  O_{\delta G_2} &= \bar{c}_v \{(i D)^2, \sigma \cdot G\} c_v \, ,
  \label{eqn:nonRPI}
\end{align}
Besides, for charm decays the power corrections in $m_s/m_c$ are defined via the dimension-seven operators:
\begin{align}
  O_{m_s^4} & =
  m_s^4 \bar{c}_v c_v \, , &
  O_{m_s^2 \mu_\pi} &=
  m_s^2 \bar{c}_v (i D)^2 c_v \, , &
  O_{ m_s^2 \mu_G} &=
  m_s^2 \bar{c}_v \sigma \cdot G c_v \, .
  \label{eqn:opm4nonRPI}
\end{align}
The coefficients of the two-quark operators, $C^\mathrm{2q}$, are determined from eq.~\eqref{eqn:OPER}. 
One first starts with $\bra{c} 2 \Im R \ket{c}$ on the l.h.s.\ of~\eqref{eqn:OPER} where the $c\to c$ transition is mediated by the effective Hamiltonian, and then divides the computation into large and small region as discussed in the previous section. This sets up the $m_s/m_c$ expansion as well as the HQE.
The result must then be  subtracted of the second term in~\eqref{eqn:OPER}, i.e.\ the renormalized one-loop matrix elements of four-quark operators multiplied by the $C^\mathrm{4q}$ found in section~\ref{sec:4qmatch} (see figure~\ref{fig:matching2q}).

For the total rate and the uncut moments, the first term in figure~\ref{fig:matching2q} is most conveniently obtained by considering the relative two-loop amplitude for $\bra{c} 2 \Im R \ket{c}$, applying the expansion~\eqref{eqn:SBGFexp} for the strange propagators and then taking the imaginary part. 
Indeed the two-loop amplitudes, which we reduced to master integrals using \texttt{FIRE6}~\cite{Smirnov:2019qkx}, depend just on one master integral~\cite{Gehrmann:1999as}:
\begin{equation}
  \int \frac{d^d q_1}{(2\pi)^d} \frac{d^d q_2}{(2\pi)^d}
  \frac{1}{q_1^2 q_2^2 (p +q_1+q_2)^2}
  = \frac{(-p^2 - i \varepsilon)^{d-3}}{(4\pi)^d (4-d)(3-d)}
  \frac{\Gamma(5-d)\Gamma^3(d/2-1)}{\Gamma(3d/2-3)} \,.
\end{equation}

The computation of the small region is easier as it reduces to one-loop diagrams. However, in the second diagram in figure~\ref{fig:matching2q} the leptonic tensor must be computed in $d$ dimension. For the total rate ($n=0$) and the $n$-th $q^2$ moment we have:
\begin{align}
  \mathcal{L}^{\mu\nu} (q) &= 
  \frac{1}{12 \pi} 
  \left( \frac{q^2}{m_c^2} \right)^n
  \left( \frac{q^2}{\mu^2} \right)^{-\varepsilon}
  \left( q^\mu q^\nu-g^{\mu\nu}q^2 \right) 
  \left[  1+  \frac{5}{3} \varepsilon \right].
\end{align}
Since the leptonic tensor is contracted with the hadronic part in eq.~\eqref{eqn:Gammasmallint} and multiplied by the result of the one-loop diagram, the $1/\varepsilon$ pole from the loop picks up the term of order $\varepsilon^1$ in $\mathcal{L}$ and gives rise to a finite difference between the second and the third term in figure~\ref{fig:matching2q}. 
Up to first order in $t$ the leptonic tensor is
\begin{align}
    \mathcal{L}^{\mu\nu} (q) &= 
    \frac{m_c^2}{12 \pi}
      \left[  1 + \varepsilon \left(\log \left(\frac{\mu^2}{m_c^2} \right)  +\frac{5}{3} \right)\right] \notag \\
    &\times
    \left[ v^\mu v^\nu - g^{\mu\nu} + \frac{1}{m_c}
      \Big( v^\mu t^\nu+t^\mu v^\nu-2 (1+n-\varepsilon) v\cdot t g^{\mu\nu}+2(n-\varepsilon) v \cdot t v^\mu v^\nu \Big)
    \right] \, ,
\end{align}
the terms of order $\varepsilon^0$ reproduces correctly the matching for the four-quark operators found in the previous section. This guarantees that the IR poles like $\log (m_s)$ cancel out in the matching. However the part in $\mathcal{L}$ proportional to $\varepsilon^1$ gives a finite contribution which is reabsorbed into $C^\mathrm{2q}$. This cancellation can be seen explicitly by considering, for instance, in the matching of $C_{\rho_D}$ for the total rate:
\begin{align}
  C_{\rho_D}(\mu) \langle O_{\rho_D} \rangle &= 
  \left[
    \frac{58}{3}+\frac{8}{\varepsilon} + 16 \log\left( \frac{\mu^2}{m_c^2} \right) 
  \right]
  \langle O_{\rho_D} \rangle   \notag  \\
  & \quad +\left[ 
  1 
  + \varepsilon \left( \frac{5}{3} -\log\left( \frac{\mu^2}{m_c^2} \right) \right) 
  \right]
  \left[ 
  -\frac{8}{\varepsilon} - 8 \log\left( \frac{\mu^2}{m_s^2} \right)+\frac{16}{3}
  +f(m_s,r_1) \right]
  \langle O_{\rho_D} \rangle \notag \\
  &\quad -
  \left[ 
   - 8 \log\left( \frac{\mu^2}{m_s^2} \right)+\frac{16}{3}
  +f(m_s,r_1) \right]
  \langle O_{\rho_D} \rangle \notag \\
  &=\left[ 6 + 8 \log\left( \frac{\mu^2}{m_c^2}  \right) \right] \langle O_{\rho_D} \rangle
  \label{eqn:rhoDexample}
\end{align}
where $\langle O_{\rho_D} \rangle = \bra{c g} O_{\rho_D} \ket{c}$ and $f(m_s,r_1)$ denotes the finite part from the loop depending on $m_s$ and  the gluon momentum $r_1$. 
The first, second and third term in~\eqref{eqn:rhoDexample}  are the contributions from the large region, the small region and the renormalized one-loop matrix element of the four-quark operator, respectively. Setting the matching scale at $\mu=m_c$ we obtain $C_{\rho_D} = 6$. 

Finally, we can compare with the expression for the $b \to c \ell \nu$ decay, which falls into case I (see the discussion in section~\ref{sec:HQEcharm}). Taking only the $\rho_D$ part for comparison, we find:
\begin{equation}
   \frac{\Gamma(B \to X_c \ell \nu)}{\Gamma_0} \Bigg|_{\rho_D} = 
   \frac{\rho_D}{m_b^3} \left[ 
   \frac{34}{3}
   +8 \log \left( \frac{m_c^2}{m_b^2} \right)
   +\mathcal{O} \left( \frac{m_c^2}{m_b^2} \right)
 \right] \, ,
 \label{eqn:rhoDbtoc}
\end{equation}
The constant term, which is independent of the quark masses, differs from that one in eq.~\eqref{eqn:rhoDexample}, while the coefficients of the logarithmic term are equal. 

In the matching of the two-quark operators the constant terms are different because in the OPE for $c \to s$ (but also in the $b \to u$ case) part of these independent contributions are reabsorbed into the matrix element of four-quark operators. 
These finite shifts arise if the four-quark operators are defined as in eqs.~\eqref{eqn:4qdefdim6} and~\eqref{eqn:4qdefdim7}.
Defining the operator basis with fierzed fields, $O^\mathrm{4q} \sim (\bar{c}^\alpha \Gamma c^\beta)(\bar{s}^\alpha \Gamma s^\beta)$, would lead to different constant terms. In particular with fierzed four-quark operators we would obtain $C_{\rho_D}=34/4$, as in the $b \to c \ell \nu$ case. 
Therefore the matching conditions of two-quark operators strictly depend on the chosen basis for the four-quark ones.
This subtlety was recognized in some of the studies of the inclusive semileptonic $b \to u \ell \nu$ decays~\cite{Bigi:2005bh,Gambino:2005tp}, while in others it has been overlooked~\cite{Gambino:2007rp,Breidenbach:2008ua,Gambino:2010jz,Ligeti:2010vd}, meaning that the four-quark operators are defined without Fierz transformation, however the coefficient of $\rho_D$ is inconsistently chosen to be the same one as in the $b \to c \ell \nu$ transition. 
 
\begin{figure}[ht]
\begin{multline*}
  \tilde C^\mathrm{2q} \times \includegraphics[width=0.15\textwidth,valign=c]{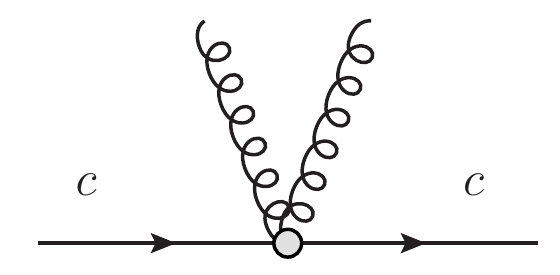} = 
   C^\mathrm{2q} \times \includegraphics[width=0.15\textwidth,valign=c]{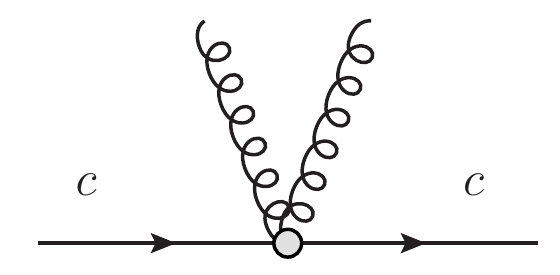}
  +C^\mathrm{4q} \times \includegraphics[width=0.2\textwidth,valign=c]{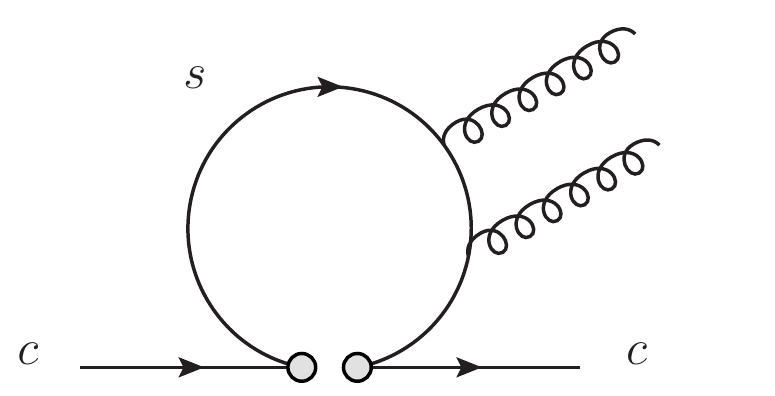}\Bigg|_\mathrm{ren} 
\end{multline*}
\caption{Second matching of two-quark operators in the OPE of case II (see section~\ref{sec:HQEcharm}).}
\label{fig:matching2qsecond} 
\end{figure}

From our expression~\eqref{eqn:rhoDexample}, we can formally recover the expression for $b \to c \ell \nu$ by evolving the Wilson coefficients from the heavy quark mass scale $m_Q$ to the light one $m_q$ (see next session) and performing a second matching at $\mu \sim m_q$, this time only onto a two-quark operator set. This procedure corresponds to case II discussed in section~\ref{sec:HQEcharm}. 
Let us call $\tilde C_{\rho_D}$ the coefficient of $O_{\rho_D}$ after the second matching. It is determined by first expanding the renormalized matrix elements of four-quark operators in the limit $m_q \gg r_i$, i.e.\ the gluon momenta are still of order $\Lambda_\mathrm{QCD}$ but the light quark mass is assumed to be a perturbative scale,  and adding the result to the matrix element of two-quark operators (see fig.~\ref{fig:matching2qsecond}). 
In our example of $\rho_D$, we would obtain:
\begin{align}
  \tilde C_{\rho_D}(m_q) &= 
  C_{\rho_D}(m_q) + \sum_i C^\mathrm{4q}_i (m_q) \langle O_i^\mathrm{4q} \rangle \notag \\
  & = \left[ 6 + 8 \log\left( \frac{m_q^2}{m_Q^2}  \right) \right] \langle O_{\rho_D} \rangle
  +
  \left[\frac{16}{3} + \mathcal{O}\left( \frac{1}{m_q^2} \right) \right] \langle O_{\rho_D} \rangle  \notag \\
  &= \left[ \frac{34}{3} + 8 \log\left( \frac{m_q^2}{m_Q^2}  \right) \right]
  \langle O_{\rho_D} \rangle,
\end{align}
which correctly reproduces the first two terms in~\eqref{eqn:rhoDbtoc}.
We explicitly verified that through this second matching procedure we can correctly reproduce the expression for $b \to c$ case up to $1/m_b^4$ for all two-quark operators.
At order $1/m_Q^5$ there are tree-level contributions to the total rate of the form $1/ m_q^2$~\cite{Bigi:2009ym,Mannel:2010wj}, which are singular in the massless limit $m_q \to 0$. 
In a two-step matching point of view, they would arise from the higher-order terms in the $1/m_q$ expansion of the four-quark matrix element.

%%%%%%%%%%%%%%%%%%%%%%%%%%%%%%%%%%%%%%%%%%%%%%%%%%
\section{Operator mixing}
\label{sec:RGE}
%%%%%%%%%%%%%%%%%%%%%%%%%%%%%%%%%%%%%%%%%%%%%%%%%%
%
The four-quark operators mix  under renormalization into the two-quark ones. 
To determine the evolution of the Wilson coefficients, we calculated  to leading terms in $\alpha_s$  the anomalous dimension matrix (ADM). 
The coefficients of the operators appearing up to order $1/m_c^4$ in the HQE can be grouped as follows:
\begin{align}
  \vec C^{2q} & = (C_{\rho_d},C_{\delta \rho_d}, C_{r_G}, C_{r_E},
  C_{s_B}, C_{s_E}, C_{s_{qB}}, C_{m_s^4},C_{m_s^2 \mu_G}) , \notag \\
  \vec C^{4q} &= (C_1, \dots, C_6) .
\end{align}
To order $\alpha_s^0$ only the coefficients in $\vec C^{2q}$ scale under renormalization according to the RGE:
\begin{align}
  \frac{\partial \vec C^{2q}}{\partial \log \mu} & = \hat \gamma^T \vec C^{4q}, 
\end{align}
while for all others we have $ \frac{\partial C_i}{\partial \log \mu} = 0$.
The ADM $\hat \gamma^T$ is obtained by computing the coefficient of the $1/\varepsilon$ pole in the one-loop matrix elements $\bra{c+ng} O_i \ket{c}$ with one or two gluons.
At the order considered, under renormalization the four-quark operators never mix into those in eq.~\eqref{eqn:nonRPI}, which do not appear in RPI observables.
This can be understood from the fact that only the RPI operators can be rewritten in term of quark states and operators in full QCD~\cite{Mannel:2018mqv}.
To leading order, the ADM is given by
\begin{equation}
  \hat \gamma^T = -\frac{1}{8 \pi^2}
  \begin{pmatrix}
    -\frac{1}{3} & \frac{2}{3} & 0 & 0 & 0 & 0 \\[3pt]
 \frac{1}{6} &  \frac{1}{3} & 0 & -\frac{1}{3} & 0 & \frac{1}{3} \\[3pt]
 \frac{1}{6} & \frac{1}{3} & \frac{1}{12} & -\frac{1}{3} & -\frac{1}{24} & \frac{1}{4} \\[3pt]
 0 & 0 & -\frac{1}{3} & 0 & \frac{1}{6} & \frac{1}{3} \\[3pt]
 \frac{1}{6} & \frac{1}{3} & -\frac{1}{4} & -\frac{1}{3} & -\frac{1}{8} & -\frac{1}{12} \\[3pt]
 0 & 0 & \frac{1}{6} & 0 & \frac{1}{12} & -\frac{1}{6} \\[3pt]
 -\frac{1}{24} & -\frac{1}{12} & \frac{1}{24} & \frac{1}{12} & \frac{1}{48} & \frac{1}{24} \\
 0 & 0 & -\frac{1}{4} & 1 & \frac{1}{8} & -\frac{3}{4} \\
 0 & 0 & -\frac{1}{2} & -1 & - \frac{1}{4} & -\frac{1}{2}
  \end{pmatrix} \, .
\end{equation}
The solution of the RGE is a simple additive logarithm:
\begin{equation}
  C^{2q}_i (\mu) = C^{2q}_i (m_c) + \log \left( \frac{\mu}{m_c} \right) 
  \sum_j \hat \gamma^T_{ij} C^{4q}_j(m_c) .
\end{equation}
Even if the renormalization group evolution is per se rather trivial at the considered order, it is instructive to compare for the total rate the logarithmic terms,
\begin{align}
  C_{\tilde \rho_D}(\mu) &=  C_{\tilde \rho_D}(m_c)+8 \log \left( \frac{\mu^2}{m_c^2} \right), & 
    C_{s_E}(\mu) &=C_{s_E}(m_c) +\frac{8}{3} \log \left( \frac{\mu^2}{m_c^2} \right), 
  \notag \\
  C_{r_G}(\mu) &=C_{r_G}(m_c) +\frac{16}{3} \log \left( \frac{\mu^2}{m_c^2} \right),&
    C_{s_{qB}}(\mu) &=C_{s_{qB}}(m_c)  - \frac{1}{3} \log \left( \frac{\mu^2}{m_c^2} \right) ,  
  \notag \\
  C_{r_E}(\mu) &=C_{r_E}(m_c) -\frac{16}{3} \log \left( \frac{\mu^2}{m_c^2} \right), &
  C_{m_s^4}(\mu) &= C_{m_s^4}(m_c) -12 \log \left( \frac{\mu^2}{m_c^2} \right), \notag \\
  C_{s_B}(\mu) &=C_{s_B}(m_c) . &
\end{align}
with the expression for the $b\to c $ case:
\begin{align}
  \frac{\Gamma(B \to X_c \ell \nu)}{\Gamma_0} &= \mu_3\left(1-8 \rho -12 \rho^2 \log \rho   \right)
  -2  \frac{\mu_G^2}{m_b^2}
  +\left( \frac{34}{3}+8 \log \rho \right) \frac{\tilde \rho_D^3}{m_b^3}
  \notag \\
  &  
   +\left( \frac{64}{9}+\frac{16}{3}\log \rho \right) \frac{r_G^4}{m_b^4}
   -\left( \frac{16}{9}+\frac{16}{3} \log \rho \right) \frac{r_E^4}{m_b^4} 
   -\frac{2}{3} \frac{s_B^4}{m_b^4}
    +\left( \frac{50}{9} +\frac{8}{3} \log \rho \right) \frac{s_E^4}{m_b^4}
   \notag \\
   & 
    -\left( \frac{25}{36} +\frac{1}{3} \log \rho \right) \frac{s_{qB}^4}{m_b^4}
    + \mathcal{O}\left( 
    \rho^3,
    \rho^2\frac{\Lambda_\mathrm{QCD}^2}{m_b^2},
    \rho\frac{\Lambda_\mathrm{QCD}^3}{m_b^3},
    \rho\frac{\Lambda_\mathrm{QCD}^4}{m_b^4}
    \right) \, ,
\end{align}
with $\rho=m_c^2/m_b^2$ and the HQE elements are defined by taking the forward-matrix element between $B$ meson states (see \cite{Fael:2018vsp, Mannel:2018mqv}).
By comparing the two expressions we see that the $\log \rho$ terms in $b \to c \ell \nu$ (case I in section~\ref{sec:HQEcharm}) are in one to one correspondence with the $\log (\mu/m_Q)$ generated by the renormalization group evolution for the $c \to s \ell \nu$ decay, which falls in case III (also for $b \to u \ell \nu$).
Similarly, we correctly reproduced these logarithms in the expressions for the  $q^2$-moments and the charged lepton energy moments.
The inclusion in the ADM of higher order corrections in $\alpha_s$ would allow us to resum term of the form $\alpha_s^n \log^{n+1}(\mu/m_Q)$. 
The phase space logarithms were resummed in~\cite{Bauer:1996ma}. 

On the contrary, the power-like singularity $1/m_q^2$ that appears at order $1/m_Q^5$ in the total rate at tree-level are not generated via the RGE mechanism.
However, as explained at the end of section~\ref{sec:2qmatch}, they arise from the second matching in case II, once the matrix element of four-quark operators is further expanded in the limit $r_i \ll m_q$.

%%%%%%%%%%%%%%%%%%%%%%%%%%%%%%%%%%%%%%%%%%%%%%%%%%
\section{A new set of HQE parameters}\label{sec:newHQE}
%%%%%%%%%%%%%%%%%%%%%%%%%%%%%%%%%%%%%%%%%%%%%%%%%%
Using the OPE described above, we obtain expressions for the total rate and $q^2$ and energy moments. They are given in Appendix~\ref{sec:app}. For the two-quark operators in eq.~\eqref{eqn:opm4}, we define the hadronic matrix elements \cite{Mannel:2018mqv,Fael:2018vsp}: 
\begin{equation}
  2 M_D X \equiv \bra{D} O_X^{2q} \ket{D},
\end{equation}
while the matrix elements of the four-quark operators given in eqs.~\eqref{eqn:4qdefdim6} and \eqref{eqn:4qdefdim7} are
\begin{equation}
  2 M_D T_i(\mu) \equiv \bra{D} O^{4q}_i \ket{D}, \quad \mbox{with } i=1,\dots,6.
\end{equation}
The four-quark matrix elements always appear in the three combinations defined in eq.~\eqref{eqn:defO0ne}, together with their $\log( \mu^2/m_c^2)$ counterpart  associated with the two-quark matrix elements. It is therefore convenient to define the parameters:
\begin{align}
  \tau_0 &= 128 \pi^2 \left(T_1-T_2 -2 \frac{T_3}{m_c}+\frac{T_4}{m_c} \right) \notag \\
  & \quad +\log \left( \frac{\mu^2}{m_c^2} \right)
  \left[ 
  8 \tilde \rho_D^3
  + \frac{1}{m_c} 
  \left( 
  \frac{16}{3} r_G^4
  -\frac{16}{3} r_E^4
  +\frac{8}{3} s_E^4
  -\frac{1}{3} s_{qB}^4
  -12 m_s^4 
  \right)
  \right], \\[5pt]
  \tau_m &= 128 \pi^2 \left(T_5-T_3 \right) \notag \\
  & \quad +\log \left( \frac{\mu^2}{m_c^2} \right)
  \left( 
  r_G^4
  -4 r_E^4
  -s_B^4
  +\frac{2}{3} s_E^4
  +\frac{1}{6} s_{qB}^4
  -3 m_s^4
  -2 m_s^2 \mu_G^2
  \right)\, , \\[5pt]
  \tau_{\epsilon} &= 64 \pi^2 \, T_6
  +\log \left( \frac{\mu^2}{m_c^2} \right)
  \left( 
  \frac{1}{3} s_B^4
  +\frac{2}{3} s_E^4
  -r_G^4
  -\frac{4}{3} r_E^4
  -\frac{1}{6} s_{qB}^4
  -\frac{4}{3}\delta\rho_D^4
  +3 m_s^4
  +2 m_s^2 \mu_G^2
  \right). 
\end{align}

We emphasize that total rate only depends on $\tau_0$, while the RPI $q^2$ moments additionally depend on $\tau_m$. As was pointed out in~\cite{Mannel:2018mqv,Fael:2018vsp}, the $q^2$ moments have the advantage that they depend on a reduced set of $10$ operators, even when including the terms up to $1/m_c^4$. For the non-RPI energy moments, additional matrix elements (up to $16$) have to be introduced. The values of the two- and four-quark matrix elements should be obtained from semileptonic charm data. Due to the large reduction of parameters, $q^2$ moments are to be preferred. However, in principle, also a combination of energy and $q^2$ moments could be used to extract the parameters. The size of the extracted coefficients would then indicate whether our OPE for semileptonic charm decays works. Finally, the obtained matrix elements should be compared to those obtained from $B$ decays. However, in the next sections we point out some subtleties concerning the extraction of the matrix elements. In addition, as discussed in Sec.~\ref{sec:matching}, the second matching step gives a finite contribution due to our basis choice of four-quark operators. In principle, these finite terms can then be reabsorbed into the $\tau_{0, m, \epsilon}$ parameters. This would then alter the coefficients of the matrix elements in the total rate and spectral moments (in which case they would match the $b\to c$ case). Of course, such a procedure would change  the numerical values obtained for the respective $\tau_i$ elements. We emphasize therefore again the importance of a consistent treatment of the four-quark operators as detailed in this paper. 

The four-quark contributions $T_i$ are usually referred to as weak-annihilation (WA) operators. Specifically, here we discussed non-valence WA since we study the weak $c\to s$ transition which does not depend on the spectator quark and is thus roughly equal for $D^+$ and $D^0$ decays. A similar argument holds for $D_s$ decays, bearing in mind that the spectator does play a role in the hadronisation such that there will be $SU(3)$ breaking effects that render the $T_i$ different for $D_s$ and $D^+$. For simplicity, we further ignore such possible $SU(3)$ breaking effects. Besides these non-valence $T_i$ contributions, also valence $T_i^{q, \rm val}$ contributions play a role. They can be obtained through a similar analysis, and by replacing $s\to q$ in the four-quark operators $O_i$ in eqs.~\eqref{eqn:4qdefdim6} and \eqref{eqn:4qdefdim7}, where $q$ is the valence (spectator) quark. The corresponding $\tau_i^{q,\rm val}$ are obtained by replacing $T_i \to T_i^{q, \rm val}$. For $D_0$ decays only non-valence operators contribute. However, for $D_s$ and $D^+$ both   valence $T_i$ and non-valence $T_i^{q, \rm val}$ contribute, with a relative weight depending on the appropriate CKM factors. The corresponding expression for the total rate and the spectral moments can be obtained by replacing:
\begin{align}
 \bar{D}^0 &: \tau_i \nonumber\\
  D_s &: \tau_i \to \tau_i + \tau_i^{s, \rm val}  \nonumber \\
D^+ &: \tau_i \to \tau_i + \left(\frac{|V_{cd}|}{|V_{cs}|}\right)^2 \tau_i^{d, \rm val} \ .
\end{align}
The valence and non-valence contributions can then be separated by taking the difference between $D_q$ and $D_0$ (see also \cite{Bigi:2009ym}). We note that the valence weak annihilation contributions are therefore in part responsible for the lifetime differences between the $D_s/D_0$ and $D^+/D_0$ in eq.~\eqref{eqn:lifedif}.  
As stated early, our results for the $c\to s$ weak transition (case III) can trivially be adapted to the $c\to d$ (case IV) by taking the limit $m_s \to 0$. Note however, the obvious change in CKM elements, explicitly:  
\begin{align}
 \bar{D}^0 &: \tau_i \nonumber\\
  D_s &: \tau_i \to \tau_i + \left(\frac{|V_{cs}|}{|V_{cd}|}\right)^2 \tau_i^{s, \rm val} \nonumber \\
 D^+ &: \tau_i \to \tau_i +  \tau_i^{d, \rm val} \ .
\end{align}
Experimentally however, it is challenging to distinguish the flavor of the light-$X$ final state and separate the $c\to s$ and $c\to d$ transitions. 

Finally, we stress that \cite{Gambino:2010jz} already used semileptonic $D$ meson data from CLEO \cite{Asner:2009pu} to extract both the valence and non-valence weak annihilation contribution of order $1/m_c^3$. However, their set-up differs from our OPE with three-expansion parameters and in the definition of the four-quark operators. In that way, the connection between the logarithmic terms accompanying $\rho_D^3$ and the weak annihilation operators is much less clear, requiring to pick a so called weak-annihilation scale. Redoing the analysis with more data and the so far unavailable $q^2$ moments, would therefore be beneficial. 

%%%%%%%%%%%%%%%%%%%%%%%%%%%%%%%%%%%%%%%%%%%%%%%%%%%
\section{Charm-Quark Mass and QCD Corrections}
\label{sec:QCD}
%%%%%%%%%%%%%%%%%%%%%%%%%%%%%%%%%%%%%%%%%%%%%%%%%%%
The HQE has a strong dependence on $m_Q$, therefore, in order to obtain precise predictions for decay rates the quark mass has to be carefully chosen. This choice is closely intertwined with the size of the QCD corrections to the decay rates. 
  
  Although all perturbative calculations are performed using the pole mass $m_Q^{\rm Pole}$, this mass 
  definition is not a good choice due to a renormalon ambiguity. This manifests itself through a bad 
  behaviour of the perturbative series that relates the pole mass to short distance 
  mass such as e.g. the $\overline{\rm MS}$ mass. However, we point out that also in the $\overline{\rm MS}$ mass scheme, the $\alpha_s$ corrections to the semileptonic decay rates have a bad convergence as well if one uses the normalization point $\mu=m_Q$ \cite{Pak:2008cp, Melnikov:2008qs,Czarnecki:1994pu,Aquila:2005hq,Melnikov:2000qh}. 
  The reason is that $\mu=m_Q$ is a poor choice since the typical energy released in inclusive decays is of the order $m_Q/5$ rather than $m_Q$. On the other hand, at very low scales the logarithmic running of the  $\overline{\rm MS}$ mass is considered unphysical \cite{Melnikov:2000qh}.
  
  Thus in order to render the QCD corrections small, an appropriate short distance mass needs to be
  chosen. One possible choice, which has been tailored for the HQE and is commonly used in semileptonic $B$ decays, is the kinetic mass. This scheme uses a hard ``Wilsonian'' cut off. It is given by \cite{Bigi:1994ga} 
  \begin{equation} 
  	m_Q^{\rm kin} (\mu) = m_Q^{\rm Pole} - [\bar\Lambda (\mu)]_{\rm pert} - \frac{1}{2 m_Q^{\rm kin} (\mu)} 
  	[ \mu_\pi^2 (\mu) ]_{\rm pert} 
	+ \mathcal{O} \left( \frac{1}{(m_Q^{\mathrm{kin}})^2} \right)\, , 
  \end{equation} 
  where the leading-order expression for 
  $\bar\Lambda$ and $\mu_\pi^2$ read \cite{Bigi:1994ga} 
  \begin{eqnarray}
  	&&  [\bar\Lambda (\mu)]_{\rm pert}  = \frac{16}{9} \frac{\alpha_s   }{\pi}  \mu   + \mathcal{O} (\alpha_s^2) \, , \\ 
  	&&   [\mu_\pi^2 (\mu)]_{\rm pert}  = \frac{4}{3} \frac{\alpha_s   }{\pi}  \mu^2  + \mathcal{O} (\alpha_s^2)  \, , 
  \end{eqnarray} 
  and $\mu$ is the cut-off scale. 
  
  Switching to the kinetic scheme, the perturbative coefficients computed in the pole scheme are  
  modified by $[\bar\Lambda (\mu)]_{\rm pert}$ and  $[ \mu_\pi^2 (\mu) ]_{\rm pert}$. 
  In addition, the parameters in the HQE are also redefined (see also \cite{Gambino:2007rp,Gambino:2004qm}):
  \begin{eqnarray}
  	\mu_\pi^2 &=& \mu_\pi^{2,{\rm kin}} + [\mu_\pi^2 (\mu)]_{\rm pert} 
  	=  \mu_\pi^{2, \rm kin} + \frac{4}{3} \frac{\alpha_s   }{\pi}  \mu^2 \, , \\ 
  	\rho_D^3  &=&\rho_D^{3,{\rm kin}}+ [\rho_D^3 (\mu)]_{\rm pert} 
  	=  \rho_D^{3,{\rm kin}} + \frac{8}{9} \frac{\alpha_s   }{\pi}  \mu^3 \, . 
\end{eqnarray} 
Therefore, terms of order $n$ in the HQE generate corrections of the order $(\alpha_s / \pi) \mu^n / m_Q^n$. This makes the choice of the cut-off scale somewhat subtle. On the one hand it 
  has to be small compared to the heavy quark mass, such that $\mu / m_Q$ is a small parameter and the $1/m_Q$ expansion remains intact, on the other hand it should be a perturbative scale.  
  
  The kinetic scheme has been successfully applied to semileptonic $B$ decays, using a cut-off scale of 1 GeV (see e.g. \cite{Alberti:2014yda, Gambino:2016jkc}), which satisfies the above criteria and leads to a highly predictive framework for inclusive $B$ decays.   
  
  For charm decays, the window for $\mu$ is much smaller, if it exists at all. The choice of 
  $\mu \sim 1$ GeV is problematic with respect to the HQE, since then $\mu / m_c \sim 1$, while 
  perturbation theory is still working. In \cite{Gambino:2010jz}, a kinetic mass for the charm with a scale choice $\mu \sim 0.5$ GeV is considered, but we emphasize that even such a choice raises questions on perturbation 
  theory. This issue should be addressed further, especially once a moment analysis is performed, but this is beyond the scope of the current work.

%%%%%%%%%%%%%%%%%%%%%%%%%%%%%%%%%%%%%%%%%%%%%%%%%%
\section{\boldmath Comparing the HQE Matrix Elements for $B$ and $D$}
\label{sec:BvsD}
%%%%%%%%%%%%%%%%%%%%%%%%%%%%%%%%%%%%%%%%%%%%%%%%%%
%
%
%
One of the motivations to investigate inclusive charm decays
is the possibility to extract the values of the HQE parameters. However, in the definitions we are using, these matrix elements depend in a non-trivial way on the  mass of the heavy quark. 

To relate the HQE elements for bottom and charm, we make use of the fact that in the $m_Q \to \infty$ limit
the matrix elements are independent on the heavy quark flavour. In order 
to study the mass dependence, we expand all quantities using 
\begin{eqnarray}
{\cal L}_{\rm QCD} &=&  \label{Lag} 
 \bar{h}_v (ivD) h_v  -  \frac{1}{2 m_Q} 
\bar{h}_v \fmslash{D}_{\perp}  \sum_{n=0}^N \left( \frac{- (ivD)}{2 m_Q} \right)^n  
\fmslash{D}_{\perp} h_v
\\
Q(x) &=& e^{-im_Qv\cdot x} \left[ h_v + H_v \right] = \label{Field}
e^{-im_Qv\cdot x} \left[ 1 +
\left( \frac{1}{2m_Q + ivD} \right) i \fmslash{D}_\perp \right] h_v  \nonumber \\
\label{HarvField}
&=& e^{-im_Qv\cdot x} \left[ 1 + \frac{1}{2m_Q} (i \fmslash{D}_\perp) + 
\left(\frac{1}{2m_Q}\right)^2 (-ivD) i\fmslash{D}_\perp + \ldots \right] h_v 
\end{eqnarray}
where $h_v$ is the static field of the heavy quark and the covariant derivative is split into a spatial and time derivative part via $iD^\mu = v^\mu\; ivD + iD^\mu_\perp$.

We start from a general matrix element 
\begin{equation} 
\langle \bar{Q}_v {\cal D} Q_v \rangle = \langle H(v) | \bar{Q}_v {\cal D} Q_v | H(v) \rangle
\end{equation}
where $H(v)$ is the heavy meson ground state and ${\cal D}$ is some combination of QCD covariant 
derivatives and $Q_v (x) = e^{im_Q v\cdot x}Q(x) $ with $Q_v = Q_v(0)$. This matrix element is defined in full QCD and depends on the heavy-quark mass (and thus they will be different for $Q=b$ and $Q=c$). Expanding it using (\ref{Lag}) and (\ref{Field}), gives
\begin{equation} \label{expand}
 \langle \bar{Q}_v {\cal D} Q_v \rangle = 
\langle \tilde{H}(v) | \bar{h}_v  {\cal D} h_v | \tilde{H}(v) \rangle  \\ 
 + \frac{1}{m_Q} \langle \tilde{H}(v) | {\cal O}_{1/m_Q}^{({\cal D})}  
| \tilde{H}(v) \rangle + \mathcal{O}(1/m_Q^2)
\end{equation} 
with 
\begin{equation}
{\cal O}_{1/m_Q}^{({\cal D})}  =  \bar{h}_v \left\{ {\cal D}, (i \fmslash{D}_\perp) \right\} h_v 
+ \frac{1}{2}\int d^4 x \, 
T \left\{ \bar{h}_v(x) (i \fmslash{D}_{\perp} )^2  h_v (x), 
 \bar{h}_v(0) {\cal D} h_v(0) \right\}
\end{equation}
where $\tilde{H}(v)$ is the heavy meson ground state in the infinite-mass limit.
We note that the first term is the correction to the operators, while the second 
term is the correction to the state. 

We emphasize that the matrix elements on the right hand side of (\ref{expand})
are independent of the heavy 
quark mass, which allows us to write 
\begin{equation}
\frac{\langle D(v) | \bar{c}_v {\cal D} c_v | D(v) \rangle}
   {\langle B(v) | \bar{b}_v{\cal D} b_v | B(v) \rangle}
= 1 + 
\left(\frac{1}{m_c}- \frac{1}{m_b} \right) 
\frac{\langle \tilde{H}(v) | {\cal O}_{1/m}^{({\cal D}) } 
| \tilde{H}(v) \rangle }{\langle \tilde{H}(v) | \bar{h}_v  {\cal D} h_v | \tilde{H}(v) \rangle }
+ \cdots 
\end{equation}  
The ratio on the right-hand side is of order $\Lambda_{\rm QCD}$ and thus the leading term is of 
order  $\Lambda_{\rm QCD} / m_c$ which can be as large as 30\%.   

An exception to this is the leading term, which can be written as \cite{Mannel:2018mqv} 
\begin{equation}
 \langle \bar{Q}_v Q_v \rangle =  \langle \bar{Q} Q \rangle 
 = \langle \bar{Q} \fmslash{v} Q \rangle + \frac{1}{2m_Q^2} \langle \bar{Q} (i \fmslash{D})^2 Q \rangle   \equiv 2 M_H \mu_3
\end{equation}
which is an exact relation in full QCD. Note that $ \langle \bar{Q} \fmslash{v} Q \rangle = 2 M_H$ and hence 
the matrix element $ \langle \bar{Q} Q \rangle$ is normalized up to terms of order $1/m_Q^2$. Following the results of \cite{Mannel:2018mqv}, we find 
\begin{equation}
\frac{2M_B}{2M_D}\frac{ \langle D(v) | \bar{c} c |D(v)  \rangle }{\langle B(v)| \bar{b} b | B(v) \rangle} = \frac{m_b}{m_c} 
\left( \frac{M_D - \bar{\Lambda}}{M_B - \bar{\Lambda}} \right) \sim 1.054 \ ,
\end{equation} 
where we used lattice input for $\bar{\Lambda} = \lim_{m_Q \to \infty} (M_H - m_Q) =0.552$ GeV~\cite{Gambino:2017vkx} and $m_b/m_c=3.78$ in the kinetic scheme at $1$ GeV \cite{Gambino:2017vkx}. Therefore the connection between the HQE elements in $B$ and $D$ decays is far from trivial, but can be quantified well enough in order to compare observables from both decays. We also emphasize, that there has been some progress on calculating the other HQE elements on the lattice in the infinite-mass limit\cite{Gambino:2017vkx}. The comparison between these determinations and those obtained from charm in the future also deserves further investigation.

%%%%%%%%%%%%%%%%%%%%%%%%%%%%%%%%%%%%%%%%%%%%%%%%%%
\section{Discussion and Conclusions}
%%%%%%%%%%%%%%%%%%%%%%%%%%%%%%%%%%%%%%%%%%%%%%%%%%

Charm physics will become an increasingly interesting field of research in the coming
few years, since - in addition to BESIII - the dedicated $B$ physics experiments 
LHCb and Belle II will collect an enormous amount of charm hadrons. Moreover, two new Super Tau-Charm Factories have been proposed at BINP, Novosibirsk, and USTC, Hefei. 

One of the most developed methods is the HQE in its application to inclusive semileptonic $b \to c$ transitions. 
On the other hand, the theoretical machinery is far from being as well developed as for bottom physics, mainly because the mass of the charm quark is between the heavy and the light quark case. While the charm quark is clearly too heavy 
to be treated in chiral perturbation theory, it remains to be explored to what extend HQE methods can be employed in charm decays. Nevertheless, there are indications that HQE methods are indeed applicable to charm decays. 

In the present paper we adapted the HQE to the case of inclusive semileptonic charm decays. We set
up this expansion for the $c\to s$ transition by treating $m_s / m_c$ in the same way as $\Lambda_{\rm QCD}/m_c$, assuming both parameters to be of the same size. Our triple expansion allowed us to systematically show how the four-quark operators are connected to the two-quark operators via renormalization. Finally, we derived the total rate and spectral moments up to $1/m_c^4$, and defined three new parameters that contain four-quark (weak annihilation) operators. We emphasize, that RPI observables, such as the total rate and $q^2$ moments, depend on a reduced set of HQE parameters \cite{Fael:2018vsp}. Therefore, it may be useful to do an experimental analysis using $q^2$ moments only, as this significantly reduces the number of free parameters.   

Dedicated experimental analyses should then answer the key question: whether the data for the total rate and the spectral moments are well described by the framework developed in this paper. Moreover, it can then be tested if the extracted HQE parameters are compatible with those extracted from $B$ decays. This would then finally show if the HQE is indeed applicable to inclusive charm decays. 

This study sets a first step towards a more systematic study of the HQE in charm decays. Making optimal use of the wealth of experimental data requires going to higher precision, and to settle the discussion on the QCD corrections and the choice of the charm mass. Eventually, this might lead to precision charm physics and towards an extraction of $V_{cs}$ and $V_{cd}$ from inclusive semileptonic charm decays.

\acknowledgments

We thank J.\ Piclum for enlightening discussions  about the method of regions.  This  work  is  supported  by  the  Deutsche  Forschungsgemeinschaft (DFG, German Research Foundation) under grant 396021762 - TRR 257 ``Particle Physics Phenomenology after the Higgs Discovery''. M.F.\ and T.M.\ thank also the Institute for Nuclear Theory at the University of Washington for its kind hospitality and stimulating research environment. This research was supported in part by the INT's U.S.\ Department of Energy grant No.\ DE-FG02-00ER41132.

\appendix
\section{Total Rate and spectral moments}\label{sec:app}
The total rate for the inclusive semileptonic charm decays can be written as
\begin{align}
    \frac{\Gamma(D \to X_s \ell \nu)}{\Gamma_0}  &= 
  \left( 1-8 \rho - 10 \rho^2 \right) \mu_3
  +\left(-2 - 8 \rho \right) \frac{\mu_G^2}{m_c^2}
  +6 \frac{\tilde \rho_D^3}{m_c^3} \notag \\
  &+\frac{16}{9}\frac{r_G^4}{m_c^4}
  +\frac{32}{9}\frac{r_E^4}{m_c^4}
  -\frac{34}{3}\frac{s_B^4}{m_c^4}
  +\frac{74}{9} \frac{s_E^4}{m_c^4}
  +\frac{47}{36}\frac{s_{qB}^4}{m_c^4}
  +\frac{\tau_0}{m_c^3}\, ,
\end{align}
with $\rho = m_s^2/m_c^2$. The rate for the decay $D \to X_d \ell \nu$ can be obtained by taking the limit $m_s \to 0$ and by making the relevant change in the four-quark operator definition inside the parameter $\tau_0$.
\subsection*{\boldmath $q^2$ moments}
\begin{align}
  \mathcal{Q}_1 {}& = 
  \left( \frac{3}{10} -\frac{9}{2} \rho- 39 \rho^2 \right) \mu_3
  +\left(-\frac{7}{5} -\frac{79}{3} \rho \right) \frac{\mu_G^2}{m_c^2}
  +\frac{41}{3} \frac{\tilde \rho_D^3}{m_c^3} \notag \\
  &\quad +\frac{527}{45}\frac{r_G^4}{m_c^4}
  -\frac{812}{45}\frac{r_E^4}{m_c^4}
  -\frac{68}{3}\frac{s_B^4}{m_c^4}
  +\frac{269}{15} \frac{s_E^4}{m_c^4}
  +\frac{111}{40}\frac{s_{qB}^4}{m_c^4}
  +\frac{\tau_0}{m_c^3}
  +2 \frac{\tau_m}{m_c^4}, \\[5pt]
  %-----------------------------------------------
  \mathcal{Q}_2 {}& = 
  \left( \frac{2}{15} -\frac{16}{5} \rho-70 \rho^2 \right) \mu_3
  -\left(\frac{16}{15} +\frac{766}{15} \rho \right) \frac{\mu_G^2}{m_c^2}
  +\frac{278}{15} \frac{\tilde \rho_D^3}{m_c^3} \notag \\
  &\quad +\frac{1013}{45}\frac{r_G^4}{m_c^4}
  -\frac{2168}{45}\frac{r_E^4}{m_c^4}
  -\frac{539}{15}\frac{s_B^4}{m_c^4}
  +\frac{1268}{45} \frac{s_E^4}{m_c^4}
  +\frac{815}{180}\frac{s_{qB}^4}{m_c^4}
  +\frac{\tau_0}{m_c^3}
  +4 \frac{\tau_m}{m_c^4},
\end{align}
  %-----------------------------------------------
\begin{align}
  \mathcal{Q}_3 {}& = 
  \left( \frac{1}{14} -\frac{5}{2} \rho-\frac{207}{2} \rho^2 \right) \mu_3
  -\left(\frac{6}{7} +\frac{397}{5} \rho \right) \frac{\mu_G^2}{m_c^2}
  +\frac{775}{35} \frac{\tilde \rho_D^3}{m_c^3} \notag \\
  &\quad +\frac{21 \, 493}{630}\frac{r_G^4}{m_c^4}
  -\frac{26 \, 378}{315}\frac{r_E^4}{m_c^4}
  -\frac{10 \, 627}{210}\frac{s_B^4}{m_c^4}
  +\frac{2459}{63} \frac{s_E^4}{m_c^4}
  +\frac{8803}{1260}\frac{s_{qB}^4}{m_c^4}
  +\frac{\tau_0}{m_c^3}
  +6 \frac{\tau_m}{m_c^4}, \\[5pt]
  %-----------------------------------------------
  \mathcal{Q}_4 {}& = 
  \left( \frac{3}{70} -\frac{72}{35} \rho-\frac{696}{5} \rho^2 \right) \mu_3
  -\left(\frac{5}{7} +\frac{11 \, 576}{105} \rho \right) \frac{\mu_G^2}{m_c^2}
  +\frac{527}{21} \frac{\tilde \rho_D^3}{m_c^3} \notag \\
  &\quad +\frac{14 \, 618}{315}\frac{r_G^4}{m_c^4}
  -\frac{38 \, 852}{315}\frac{r_E^4}{m_c^4}
  -\frac{6967}{105}\frac{s_B^4}{m_c^4}
  +\frac{5293}{105} \frac{s_E^4}{m_c^4}
  +\frac{7951}{840}\frac{s_{qB}^4}{m_c^4}
  +\frac{\tau_0}{m_c^3}
  +8 \frac{\tau_m}{m_c^4}.
\end{align}
\subsection*{\boldmath Electron energy moments}
\begin{align}
  \mathcal{Y}_1 {}& = 
  \frac{3}{5} - 6 \rho- 23 \rho^2 
  -\left(1 + 16 \rho \right) \frac{\mu_G^2}{m_c^2}
  +\frac{139}{15} \frac{\rho_D^3}{m_c^3} 
  +\frac{3}{5} \frac{\rho_{LS}^3}{m_c^3} 
  +\frac{503}{90} \frac{\delta \rho_D^3}{m_c^3} 
  +\frac{3}{10} \frac{\delta \rho_{LS}^3}{m_c^3} 
  \notag \\
  &\quad +\frac{1271}{180}\frac{r_G^4}{m_c^4}
  -\frac{208}{45}\frac{r_E^4}{m_c^4}
  -\frac{682}{45}\frac{s_B^4}{m_c^4}
  +\frac{203}{15} \frac{s_E^4}{m_c^4}
  +\frac{283}{180}\frac{s_{qB}^4}{m_c^4}
  +\frac{1}{4} \frac{\delta_{G2}^4}{m_c^4}
  +\frac{\tau_0}{m_c^3}
  + \frac{\tau_m}{m_c^4}
  +\frac{\tau_\epsilon}{m_c^4}, \\[5pt]
  %-----------------------------------------------
  \mathcal{Y}_2 {}& = 
  \frac{2}{5} - \frac{24}{5} \rho - 35 \rho^2 
  +\left(\frac{1}{3} - 4 \rho \right) \frac{\mu_\pi^2}{m_c^2}
  -\left(\frac{11}{15} + 20 \rho \right) \frac{\mu_G^2}{m_c^2}
  +\frac{34}{3} \frac{\rho_D^3}{m_c^3} 
  +\frac{14}{15} \frac{\rho_{LS}^3}{m_c^3} 
  \notag \\
  &\quad
  +\frac{64}{9} \frac{\delta \rho_D^3}{m_c^3} 
  +\frac{2}{5} \frac{\delta \rho_{LS}^3}{m_c^3} 
  +\frac{536}{45}\frac{r_G^4}{m_c^4}
  -\frac{74}{5}\frac{r_E^4}{m_c^4}
  -\frac{872}{45}\frac{s_B^4}{m_c^4}
  +\frac{866}{45} \frac{s_E^4}{m_c^4}
  +\frac{167}{90}\frac{s_{qB}^4}{m_c^4}
  \notag \\
  &\quad
  +\frac{2}{15} \frac{\delta_{G1}^4}{m_c^4}
  +\frac{8}{15} \frac{\delta_{G2}^4}{m_c^4}
  +\frac{\tau_0}{m_c^3}
  +2\frac{\tau_m}{m_c^4}
  +2\frac{\tau_\epsilon}{m_c^4}, \\[5pt]
  %-----------------------------------------------
  \mathcal{Y}_3 {}& = 
  \frac{2}{7} - 4 \rho - \frac{232}{5} \rho^2 
  +\left(\frac{4}{7} - 8 \rho \right) \frac{\mu_\pi^2}{m_c^2}
  -\left(\frac{4}{7} + 24 \rho \right) \frac{\mu_G^2}{m_c^2}
  +\frac{446}{35} \frac{\rho_D^3}{m_c^3} 
  +\frac{8}{7} \frac{\rho_{LS}^3}{m_c^3} 
  \notag \\
  &\quad
  +8 \frac{\delta \rho_D^3}{m_c^3} 
  +\frac{2}{5} \frac{\delta \rho_{LS}^3}{m_c^3} 
  +\frac{3469}{210}\frac{r_G^4}{m_c^4}
  -\frac{394}{15}\frac{r_E^4}{m_c^4}
  -\frac{5011}{210}\frac{s_B^4}{m_c^4}
  +\frac{379}{15} \frac{s_E^4}{m_c^4}
  +\frac{43}{20}\frac{s_{qB}^4}{m_c^4}
  \notag \\
  &\quad
  +\frac{3}{7} \frac{\delta_{G1}^4}{m_c^4}
  +\frac{29}{35} \frac{\delta_{G2}^4}{m_c^4}
  +\frac{\tau_0}{m_c^3}
  +3\frac{\tau_m}{m_c^4}
  +3\frac{\tau_\epsilon}{m_c^4}, \\[5pt]
  %-----------------------------------------------
  \mathcal{Y}_4 {}& = 
  \frac{3}{14} - \frac{24}{7} \rho - \frac{287}{5} \rho^2 
  +\left(\frac{3}{4} - 12 \rho \right) \frac{\mu_\pi^2}{m_c^2}
  -\left(\frac{13}{28} + 28 \rho \right) \frac{\mu_G^2}{m_c^2}
  +\frac{481}{35} \frac{\rho_D^3}{m_c^3} 
  +\frac{9}{7} \frac{\rho_{LS}^3}{m_c^3} 
  \notag \\
  &\quad
  +\frac{178}{21} \frac{\delta \rho_D^3}{m_c^3} 
  +\frac{12}{35} \frac{\delta \rho_{LS}^3}{m_c^3} 
  +\frac{2937}{140}\frac{r_G^4}{m_c^4}
  -\frac{1352}{35}\frac{r_E^4}{m_c^4}
  -\frac{5993}{210}\frac{s_B^4}{m_c^4}
  +\frac{1104}{35} \frac{s_E^4}{m_c^4}
  +\frac{103}{42}\frac{s_{qB}^4}{m_c^4}
  \notag \\
  &\quad
  +\frac{9}{10} \frac{\delta_{G1}^4}{m_c^4}
  +\frac{79}{70} \frac{\delta_{G2}^4}{m_c^4}
  +\frac{\tau_0}{m_c^3}
  +4\frac{\tau_m}{m_c^4}
  +4\frac{\tau_\epsilon}{m_c^4}. 
\end{align}

\newpage
\label{Bibliography}
\bibliographystyle{JHEP}
\bibliography{BIB}
\end{document}